\documentclass[preprint2,tighten]{aastex6}
\pdfoutput=1 
\usepackage{amsmath,amstext}
\usepackage[T1]{fontenc}
\usepackage{apjfonts} 
\usepackage[figure,figure*]{hypcap}
\usepackage{multirow}
\usepackage[T1]{fontenc} 
\usepackage[utf8]{inputenc} 
\usepackage{float}

\newcommand{\unit}[1]{\ensuremath{\, \mathrm{#1}}}

\shorttitle{A sub-Neptune sized planet transiting the M2.5-dwarf G 9-40}
\shortauthors{Stefansson et al. 2019}

\begin{document}
\title{A sub-Neptune sized planet transiting the M2.5-dwarf G 9-40: Validation with the Habitable-zone Planet Finder}
\author{Gudmundur Stefansson\altaffilmark{1,2,3,4,5,6}}
\author{Caleb Ca\~nas\altaffilmark{1,2,3,4}}
\author{John Wisniewski\altaffilmark{7}}
\author{Paul Robertson\altaffilmark{8}}
\author{Suvrath Mahadevan\altaffilmark{1,2,3}}
\author{Marissa Maney\altaffilmark{1}}
\author{Shubham Kanodia\altaffilmark{1,2}}
\author{Corey Beard\altaffilmark{8}}
\author{Chad F. Bender\altaffilmark{9}}
\author{Peter Brunt\altaffilmark{10}}
\author{J. Christopher Clemens\altaffilmark{11}}
\author{William Cochran\altaffilmark{12}}
\author{Scott A. Diddams\altaffilmark{13,14}}
\author{Michael Endl\altaffilmark{12}}
\author{Eric B. Ford\altaffilmark{1,2,3,15}}
\author{Connor Fredrick\altaffilmark{13,14}}
\author{Samuel Halverson\altaffilmark{16,17}}
\author{Fred Hearty\altaffilmark{1,2}}
\author{Leslie Hebb\altaffilmark{18}}
\author{Joseph Huehnerhoff\altaffilmark{19}}
\author{Jeff Jennings\altaffilmark{13,14}}
\author{Kyle Kaplan\altaffilmark{9}}
\author{Eric Levi\altaffilmark{1}}
\author{Emily Lubar\altaffilmark{1}}
\author{Andrew J. Metcalf\altaffilmark{20,13,14}}
\author{Andrew Monson\altaffilmark{1}}
\author{Brett Morris\altaffilmark{19}}
\author{Joe P. Ninan\altaffilmark{1,2}}
\author{Colin Nitroy\altaffilmark{1}}
\author{Lawrence Ramsey\altaffilmark{1,2,3}}
\author{Arpita Roy\altaffilmark{21,22}}
\author{Christian Schwab\altaffilmark{23}}
\author{Steinn Sigurdsson\altaffilmark{1,2,3}}
\author{Ryan Terrien\altaffilmark{24}}
\author{Jason T. Wright\altaffilmark{1,2,3}}
\email{gstefansson@astro.princeton.edu}

\altaffiltext{1}{Department of Astronomy \& Astrophysics, The Pennsylvania State University, 525 Davey Lab, University Park, PA 16802, USA}
\altaffiltext{2}{Center for Exoplanets \& Habitable Worlds, University Park, PA 16802, USA}
\altaffiltext{3}{Penn State Astrobiology Research Center, University Park, PA 16802, USA}
\altaffiltext{4}{NASA Earth and Space Science Fellow}
\altaffiltext{5}{Department of Astrophysical Sciences, Princeton University, 4 Ivy Lane, Princeton, NJ 08540, USA}
\altaffiltext{6}{Henry Norris Russell Fellow}
\altaffiltext{7}{Homer L. Dodge Department of Physics and Astronomy, University of Oklahoma, 440 W. Brooks Street, Norman, OK 73019, USA}
\altaffiltext{8}{Department of Physics and Astronomy, University of California-Irvine, Irvine, California 92697, USA}
\altaffiltext{9}{Steward Observatory, University of Arizona, Tucson, Arizona 85721, USA}
\altaffiltext{10}{AVR Optics, 187 North Water Street, Rochester, New York, 14604, USA}
\altaffiltext{11}{University of North Carolina at Chapel Hill, Chapel Hill, NC 27599, USA}
\altaffiltext{12}{Department of Astronomy and McDonald Observatory, University of Texas at Austin, 2515 Speedway, Stop C1400, Austin, TX 78712, USA}
\altaffiltext{13}{Time and Frequency Division, National Institute of Standards and Technology, 325 Broadway, Boulder, CO 80305, USA}
\altaffiltext{14}{Department of Physics, University of Colorado, 2000 Colorado Avenue, Boulder, CO 80309, USA}
\altaffiltext{15}{Institute for CyberScience, The Pennsylvania State University, 525 Davey Laboratory, University Park, PA 16802, USA}
\altaffiltext{16}{NASA Sagan Fellow}
\altaffiltext{17}{MIT Kavli Institute for Astrophysics, 70 Vassar St, Cambridge, MA 02109, USA}
\altaffiltext{18}{Department of Physics, Hobart and William Smith Colleges, 300 Pulteney Street, Geneva, NY 14456, USA}
\altaffiltext{19}{Department of Astronomy, Box 351580, University of Washington, Seattle, WA 98195, USA}
\altaffiltext{20}{Space Vehicles Directorate, Air Force Research Laboratory, 3550 Aberdeen Ave SE, Kirtland AFB, NM 87117, USA}
\altaffiltext{21}{Millikan Prize Postdoctoral Fellow}
\altaffiltext{22}{California Institute of Technology, 1200 E California Blvd, Pasadena, California 91125, USA}
\altaffiltext{23}{Department of Physics and Astronomy, Macquarie University, Balaclava Road, North Ryde, NSW 2109, Australia}
\altaffiltext{24}{Department of Physics and Astronomy, Carleton College, Northfield, Minnesota 55057, USA}

\begin{abstract}
We validate the discovery of a 2 Earth radii sub-Neptune-size planet around the nearby high proper motion M2.5-dwarf G 9-40 (EPIC 212048748), using high-precision near-infrared (NIR) radial velocity (RV) observations with the Habitable-zone Planet Finder (HPF), precision diffuser-assisted ground-based photometry with a custom narrow-band photometric filter, and adaptive optics imaging. At a distance of $d=27.9\unit{pc}$, G 9-40b is the second closest transiting planet discovered by \textit{K2} to date. The planet's large transit depth ($\sim$3500ppm), combined with the proximity and brightness of the host star at NIR wavelengths (J=10, K=9.2) makes G 9-40b one of the most favorable sub-Neptune-sized planet orbiting an M-dwarf for transmission spectroscopy with JWST, ARIEL, and the upcoming Extremely Large Telescopes. The star is relatively inactive with a rotation period of $\sim$29 days determined from the \textit{K2} photometry. To estimate spectroscopic stellar parameters, we describe our implementation of an empirical spectral matching algorithm using the high-resolution NIR HPF spectra. Using this algorithm, we obtain an effective temperature of $T_{\mathrm{eff}}=3404\pm73$K, and metallicity of $\mathrm{[Fe/H]}=-0.08\pm0.13$. Our RVs, when coupled with the orbital parameters derived from the transit photometry, exclude planet masses above $11.7 M_\oplus$ with 99.7\% confidence assuming a circular orbit. From its radius, we predict a mass of $M=5.0^{+3.8}_{-1.9} M_\oplus$ and an RV semi-amplitude of $K=4.1^{+3.1}_{-1.6}\unit{m\:s^{-1}}$, making its mass measurable with current RV facilities. We urge further RV follow-up observations to precisely measure its mass, to enable precise transmission spectroscopic measurements in the future.
\end{abstract}
\keywords{planets and satellites: detection --- planetary systems --- stars: fundamental parameters}

\section{Introduction}
\noindent
Surveying the ecliptic plane, the \textit{K2} mission \citep{howell2014} has expanded on the exoplanet discoveries of the original \textit{Kepler} \citep{borucki2010} mission. In particular, \textit{K2} has sampled a different population of planet hosts than \textit{Kepler}, namely detecting planets orbiting brighter and closer stars, and stars of later spectral type. Among these, planets around bright, nearby M-dwarfs are compelling targets for further characterization for many reasons.

First, planets around bright M-dwarfs are amenable to precision radial velocity (RV) observations---especially for Doppler spectrographs operating in the near-infrared (NIR)---to measure their masses, and to search for additional planets. Measuring their masses will help shed light onto the exoplanet Mass-Radius relation \citep{weiss2014,wolfgang2016,chen2017,ning2018} of planets around M-dwarfs. Recent work by \cite{kanodia2019}, comparing the exoplanet Mass-Radius relation of M-dwarf planets to that of the $\textit{Kepler}$ sample of earlier-type planet hosts hints at a difference between the resulting Mass-Radius relationships. However, the comparison is still heavily influenced by the 7 TRAPPIST-1 planets, which account for $\sim$$30\%$ of known M-dwarf planets with precisely determined masses and radii. To confirm if there is a statistical difference in the Mass-Radius relationship between M-dwarf planets and earlier-type planets, we need to increase the number of transiting M-dwarf planets with precisely measured masses.

Second, the large planet-to-star radius ratios of planets orbiting M-dwarfs make them favorable targets for transmission spectroscopy with upcoming facilities, such as JWST \citep{cowan2015}, ARIEL \citep{tinetti2016} and the Extremely Large Telescopes (ELTs). A precise measurement of the planetary mass is a requirement for inference of atmospheric features to disentangle degeneracies that exist between the atmospheric scale height and the mean molecular weight of the atmosphere \citep{batalha2017}. Although recent studies \citep[e.g.,][]{crossfield2017} have demonstrated rising statistical trends that show cold ($T_{\mathrm{eq}}<800 \unit{K}$) sub-Neptune sized planets tend to have damped transmission spectroscopic features, sub-Neptune sized planets have no analogue in the Solar System, making them particularly interesting targets as they are observed to have a diverse range of compositions and observed radii \citep[e.g.,][]{fulton2017} resulting in a diverse range of possible atmospheres.

In this paper, we validate the planetary nature of a $R\sim2.0R_\oplus$ sub-Neptune sized planet orbiting the nearby high proper motion M2.5-dwarf star G 9-40, also known as EPIC 212048748, NLTT 20661, 2MASS J08585232+2104344, and Gaia DR2 684992690384102528. At a distance of $d=27.9\unit{pc}$, G 9-40b is currently the second closest planetary system discovered by the \textit{K2} mission. G 9-40b was originally identified as a planet candidate in \textit{K2} Campaign 16 by \cite{yu2018}, and here we perform the necessary observations to validate the planet candidate as a \textit{bona-fide} planet through precision ground-based diffuser-assisted transit photometry using the Astrophysical Research Consortium Telescope Imaging Camera (ARCTIC) \citep{huehnerhoff2016} on the Astrophysical Research Consortium (ARC) 3.5m Telescope at Apache Point Observatory, adaptive optics imaging using the ShaneAO instrument \citep{srinath2014} on the 3m Shane Telescope at Lick Observatory, and precision near-infrared (NIR) radial velocities obtained with the Habitable-zone Planet Finder (HPF) Spectrograph \citep{mahadevan2012,mahadevan2014}. Using the HPF NIR spectra, we provide precise spectroscopic parameters using an empirical spectral-matching technique and use the spectra for precision velocimetry to provide an upper bound on the mass of G 9-40b.

The paper is structured as follows. In Section 2 we discuss the observations used in this paper. We discuss our best estimates of the stellar parameters of the host star in Section 3, describing our implementation of an empirical spectral matching algorithm closely following the popular \texttt{SpecMatch-emp} algorithm from  \cite{yee2017}. In Section 4 we discuss the transit analysis of the \textit{K2} and ground-based data and the resulting best-fit planet parameters. In Section 5 we discuss our false positive analysis using the VESPA statistical validation tool where we statistically validate G 9-40b as a planet. In Section 6 we compare G 9-40b to other known planetary systems, and provide a further discussion of the feasibility for future study of this system through transmission spectroscopy and precision RV observations. Finally, we conclude the paper in Section 7 with a summary of our key results.

\section{Observations}
\subsection{K2 Photometry}
G 9-40 was observed by the \textit{Kepler} spacecraft as part of Campaign 16 of the \textit{K2} mission. It was proposed as a \textit{K2} Campaign 16 target by the following programs: $\mathrm{GO16005\_LC}$ (PI: Crossfield), $\mathrm{GO16009\_LC}$ (PI: Charbonneau), $\mathrm{GO16052\_LC}$ (PI: Stello), and $\mathrm{GO16083\_LC}$ (PI: Coughlin). The star was monitored in long cadence mode (30 min cadence) for 80 days from December 7, 2017 to February 25, 2018 and was originally identified as a candidate planet host by \cite{yu2018}. We used the \texttt{Everest} pipeline \citep{luger2016,luger2017} to detrend and correct for photometric variations seen in the \textit{K2} photometry due to imperfect pointing of the spacecraft with only 2 functioning reaction wheels. The unbinned 6.5 hour photometric precision before detrending was 193ppm and it decreased to 30ppm after detrending with \texttt{Everest}.

\subsection{High-resolution Doppler Spectroscopy}
\label{sec:spectra}
We obtained 4 visits of G 9-40 with the Habitable-zone Planet Finder (HPF) Spectrograph with the goal to measure its radial velocity (RV) variation as a function of time. HPF is a high-resolution ($R\sim55,000$) near-infrared (NIR) spectrograph recently commissioned on the 10m Hobby-Eberly Telescope (HET) in Texas covering the information-rich $z$, $Y$ and $J$ bands from $810-1280 \unit{nm}$ \citep{mahadevan2012,mahadevan2014}. HPF is actively temperature-stabilized achieving $\sim$1$\unit{mK}$ temperature stability long-term \citep{stefansson2016}. The HET is a fully queue-scheduled telescope with all observations executed in a queue by the HET resident astronomers \citep{shetrone2007}. In each visit, we obtained two 945s exposures, yielding a total of 8 spectra with a median Signal-to-Noise (S/N) of 123 per extracted 1D pixel at $\lambda = 1000\unit{nm}$. 

HPF has a NIR laser-frequency comb (LFC) calibrator which has been shown to enable $\sim20\unit{cm\:s^{-1}}$ calibration precision and $1.53\unit{m\:s^{-1}}$ RV precision on-sky on the bright and stable M-dwarf Barnard's Star \citep{metcalf2019}. We elected to not use the simultaneous LFC reference calibrator for these observations to minimize any possibility of scattered LFC light in the target spectrum. Instead, the drift correction was performed by extrapolating the wavelength solution from other LFC exposures from the night of the observations. As is discussed in detail in Appendix A, HPF's nightly drift is a predictable linear saw-tooth pattern with a $10-15 \unit{m\:s^{-1}}$ amplitude (see Figure \ref{fig:drift3} in Appendix A), where dedicated LFC calibration exposures are generally taken nightly a few hours apart, enabling precise wavelength calibration even without simultaneous LFC calibration exposures \citep{metcalf2019}. To further illustrate this linear behavior, in Appendix A, we show the LFC drift exposures for the four nights we obtained G 9-40 observations. Lastly, in Appendix A, we also estimate the error contribution of the drifts to the total RV errors using a leave-one-approach. In doing so, we estimate errors at the $<30 \unit{cm/s}$ level for our G 9-40 observations, which we add in quadrature to the errors estimated from our RV extraction (see Table \ref{tab:rvs}).

The HPF 1D spectra were reduced and extracted with the custom HPF data-extraction pipeline following the procedures outlined in \cite{ninan2018}, \cite{kaplan2018}, and \cite{metcalf2019}. After the 1D spectral extraction, we extracted precise NIR RVs using a modified version of the SpEctrum Radial Velocity Analyzer (SERVAL) pipeline \citep{zechmeister2018} following the methodology discussed in \cite{metcalf2019}. In short, SERVAL uses a template-matching technique to derive RVs \citep[see ][]{anglada2012,zechmeister2018}. This technique relies on minimizing the differences of the observed spectrum against a high S/N master template constructed from a co-addition of all available observations of the target star. We generated the master template by performing a best-fit spline-regression to the 8 as-observed spectra of the target star. Following the methodology described in \cite{metcalf2019}, we masked out all telluric and sky-emission line regions, using a thresholded synthetic telluric-line mask and an empirical thresholded sky-emission line mask, respectively. We generated the telluric-line mask using the \texttt{telfit} \citep{gullikson2014} Python wrapper to the Line-by-Line Radiative Transfer Model package (\texttt{LBLRTM}) \citep{clough2005}. For the sky-emission line mask, we used the same empirical blank-sky sky-emission line mask as used in \cite{metcalf2019}. We calculated the barycentric corrections using the \texttt{barycorrpy} Python package \cite{kanodia2018}, which uses the barycentric-correction algorithms from \cite{wright2014}.

\subsection{Adaptive Optics Imaging}
To constrain contamination from background sources and study the possibility that astrophysical false positives could be the source of the transits (e.g., eclipsing binaries or background eclipsing binaries), we performed high-contrast adaptive optics (AO) imaging of G 9-40 using the 3\,m Shane Telescope at Lick Observatory on November 25, 2018. We performed the AO imaging using the upgraded ShARCS camera \citep{srinath2014} in the $K_s$ bandpass. We observed the star using a 5-point dither pattern \citep[see e.g.,][]{furlan2017}, imaging the star at the center of the detector and in each quadrant. Individual images had an exposure time of 30 seconds and we obtained a total of 20 minutes integration time across all exposures.

\begin{figure}[t]
\begin{center}
\includegraphics[width=\columnwidth]{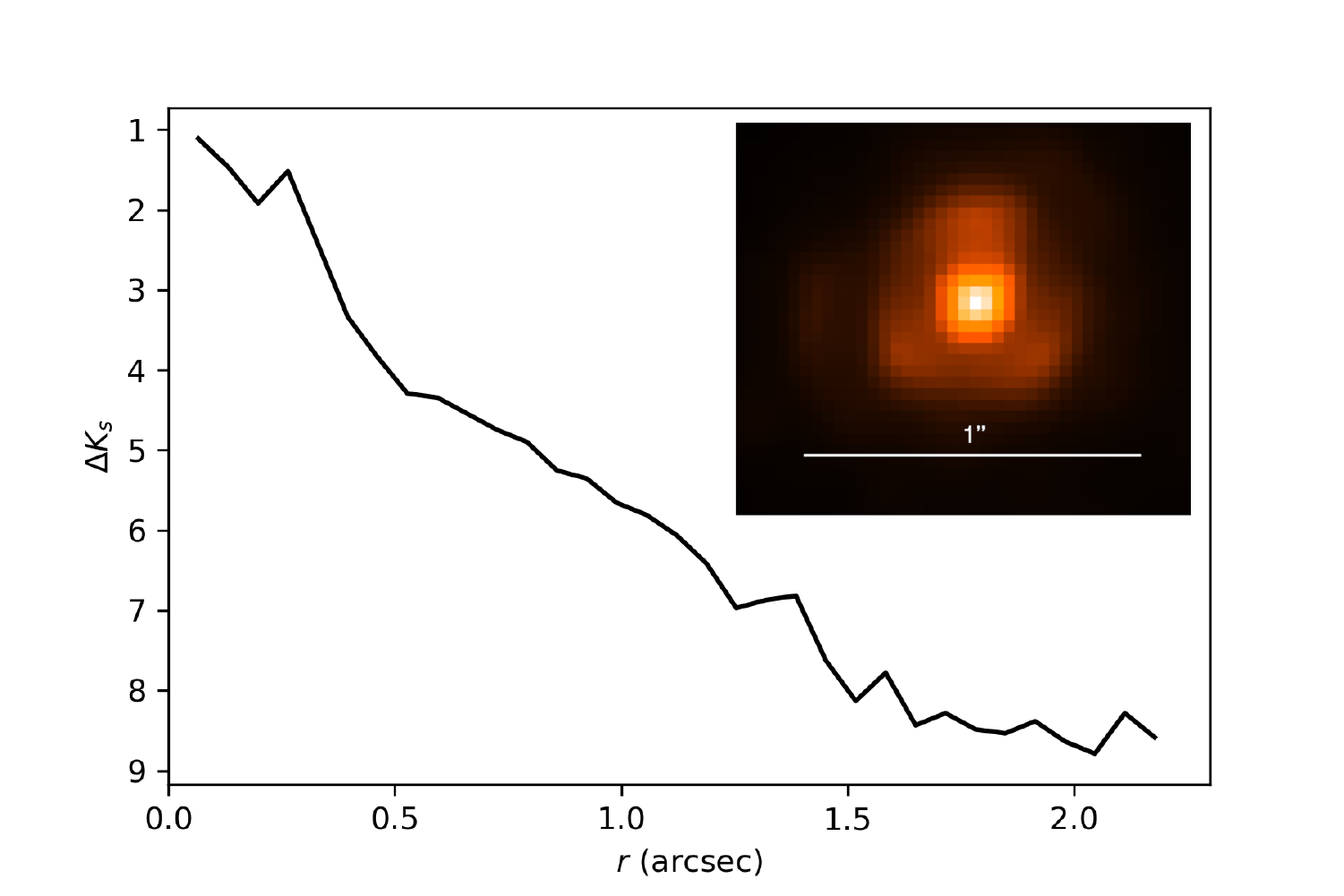}
\end{center}
\vspace{-0.3cm}
\caption{Adaptive optics (AO) $K_s$ band contrast curve (azimuthally averaged) of G 9-40 from ShARCS on the Shane 3\,m Telescope. No bright companions are seen to within $0.5\arcsec$ separation. The inset image shows the as-observed AO image along with a $1\arcsec$ bar for scale.}
\label{fig:contrast}
\end{figure}

Standard image processing (e.g., flat fielding, sky subtraction) as well as sub-pixel image alignment was performed using custom Python software. Using the combined image, we computed the variance in flux in a series of concentric annuli centered on the target star. The resulting $5\sigma$ contrast curve is shown in Figure \ref{fig:contrast}. From the images we see that no bright ($\Delta K_s$ < 4) secondary companions are detectable within $0.5\arcsec$. These data show that there is no evidence of blending up to a radius of $2.5\arcsec$. We therefore infer that the transits observed from both \textit{K2} and from the ground (see Section 4) are not contaminated by background sources.

\subsection{Seeing Limited Imaging}
\begin{figure*}[t]
\begin{center}
\includegraphics[width=0.85\textwidth]{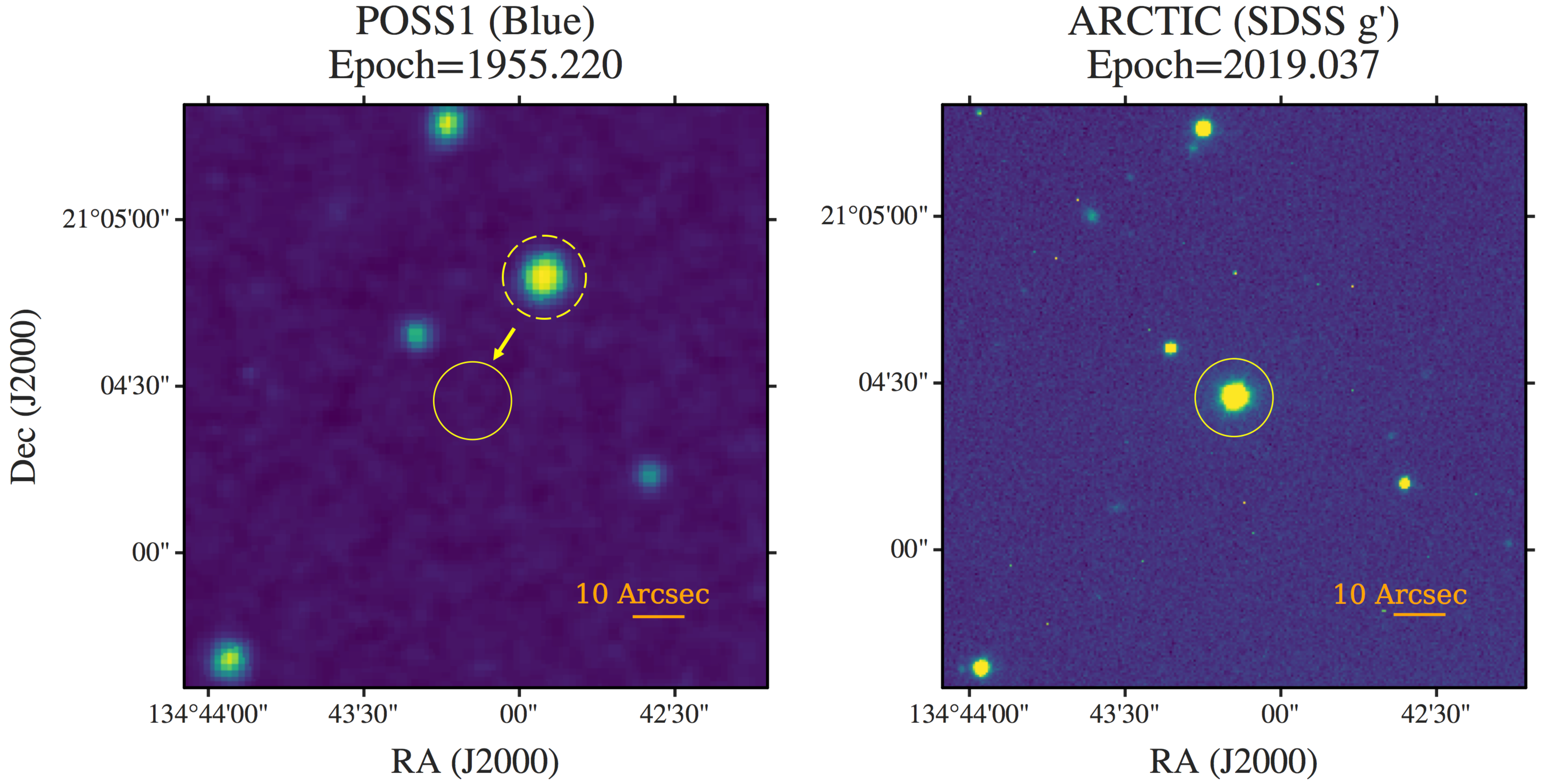}
\end{center}
\vspace{-0.3cm}
\caption{a) Seeing limited image of G 9-40 obtained by POSS-1 in the Blue filter from the Digital Sky Survey-1 (DSS-1) from 1955. b) Seeing limited image using the ARCTIC imager on the 3.5m ARC Telescope at APO in the SDSS g$^\prime$ filter from our 2019 observations. The filled yellow circle (7\arcsec radius) denotes the position of G 9-40 at epoch 2019.037, and the dashed yellow circle in a) highlights the position of G 9-40 at the epoch of the POSS1 plate (1955.22). From the POSS1 image we see that no background star is visible at the position of G 9-40 during the ARCTIC and the \textit{K2} transit observations. This suggests that G 9-40 is the true source of the transits. North is up, and East is to the left.}
\label{fig:seeing}
\end{figure*}

To constrain blends within the \textit{K2} aperture at distances further than the $2.5\arcsec$ FOV of the diffraction limited Lick/ShaneAO images shown in Figure \ref{fig:contrast}, we obtained seeing-limited images using the ARCTIC imager \citep{huehnerhoff2016} on the 3.5m Astrophysical Research Consortium Telescope (Figure \ref{fig:seeing}b). On the night of January 14 2019, we obtained 20 seeing limited images of G 9-40 using an exposure time of $10 \unit{s}$ in the SDSS g$^\prime$ filter. The skies were photometric with a seeing of $0.75\arcsec$. We median combined the full set of 20 images to create a high S/N final image shown in Figure \ref{fig:seeing}. Given G 9-40's high proper motion, we compare our median-combined seeing limited image to archival observations from the POSS1 survey obtained using the \texttt{astroquery skyview} service \citep{astroquery} in Figure \ref{fig:seeing}. The POSS1 image shows no evidence for a background star at G 9-40's current coordinates marked by the yellow circle (7$\arcsec$ radius) in both panels in Figure \ref{fig:seeing}. In addition to this, we also queried the \textit{Gaia} archive \citep{gaia2018} for nearby sources. In doing so, \textit{Gaia} reveals one companion at a distance of $20\arcsec$ from G 9-40, with a $\Delta G = 5$ which is both significantly fainter than G 9-40 and outside the \textit{K2} aperture. Given G 9-40's proper motion of $\mu_\alpha = 175 \unit{mas\:yr^{-1}}$ and $\mu_\delta = -318 \unit{mas\:yr^{-1}}$, and the short timespan between the \textit{Gaia} and \textit{K2} observations, we conclude that this star is not a significant source of dilution in the \textit{K2} and diffuser-assisted transits.

\subsection{Diffuser-assisted Photometry from the ARC 3.5m}
We observed the transit of G 9-40b using the ARCTIC imager \citep{huehnerhoff2016} on the Astrophysical Research Consortium (ARC) 3.5m Telescope at Apache Point Observatory on the night of April 13, 2019 local time (04:50UT - 07:30UT April 14). The night was photometric with a seeing of $\sim$0.75\arcsec at the beginning of the night. The target rose from airmass 1.55 to airmass 1.05 during the observations.

We observed the transit using the Engineered Diffuser on the ARCTIC imager, which has been described in detail in \cite{stefansson2017,stefansson2018a}. In short, the Engineered Diffuser molds the focal-plane image of the star into a broad and stable top-hat shape, allowing us to increase our exposure times to gather more photons per exposure while minimizing correlated errors due to point spread function (PSF) variations and guiding errors. The increased exposure time allows us to further average over scintillation errors by increasing the duty cycle of the observations \citep{stefansson2017}. We used a newly-commissioned, custom-fabricated narrow-band filter from AVR Optics\footnote{\url{http://avr-optics.com/}} and Semrock Optics\footnote{\url{https://www.semrock.com/}}, centered on 857nm with a 37nm width in a region with minimal telluric absorption. The filter, along with its as-measured throughput curve, has been further discussed in \citep{stefansson2018b}. The filter is 5 inches in diameter, covering the full ARCTIC beam-footprint in the optical path, resulting in minimal vignetting and allows the observer to make use of the full field-of-view (FoV) of ARCTIC. The combination of the Engineered Diffuser and the narrow-band filter allowed us to make use of a bright reference star HD 76780 ($V=7.63, I=6.7$; G 9-40 has a SDSS i$^\prime$ magnitude of 11.994) within the FoV at a high observing efficiency of 76.9\% with a cadence of 18.2s between successive exposures (exposure time of 14s)\footnote{To further minimize readout-time and maximize our observing efficiency, we used ARCTIC in the $2\times2$ quad-amplifier binning mode (gain of $1.97 \unit{e^{-}/ADU}$, read-noise of $6.7\unit{e^-}$) with a readout time of 4.2s.}.

We reduced the on-sky transit observations along with the 25 bias and 25 dome flat-field calibration frames using AstroImageJ \citep{collins2017} following the methodology in \cite{stefansson2017}. We calculate the Barycentric Julian Date ($\mathrm{BJD_{TDB}}$) time using the Python package \texttt{barycorrpy}. After experimenting with different aperture radii and annuli, we found that the aperture setting that yielded the smallest residuals was a value of $30\unit{pixels}$ (6.9$\arcsec$) for the target star aperture, $40\unit{pixels}$ (9.2$\arcsec$) for the inner annulus, and $80\unit{pixels}$ (18.4$\arcsec$) for the outer annulus for background estimation. This resulted in peak ADU/pixel counts of $\sim$1,000 and $\sim$42,900 for the target and reference star, respectively. After removing the best-fit transit model and additional 6 significant ($>3 \sigma$) outliers present in the data, we obtained an unbinned precision of 1225ppm which is further discussed in Section \ref{sec:transits}. The transit is shown in Figure \ref{fig:transit} where the data is shown with error bars estimated following the methodology in \cite{stefansson2017} and accounting for errors from photon, dark, read, background, digitization and scintillation noise. 

\section{Stellar Parameters}

\subsection{Stellar Parameters from Matching to an Empirical Library of Spectra}
In this subsection we discuss our implementation of the \texttt{SpecMatch-Emp} algorithm described in \cite{yee2017} to derive precise estimates of the spectroscopic effective temperature ($T_{\mathrm{eff}}$), metallicity ([Fe/H]), and surface gravity ($\log g$) of G 9-40 through comparing our high-resolution NIR HPF spectra of G 9-40 to a library of high S/N as-observed HPF spectra with well-characterized stellar parameters. As this is the first description of this implementation for this algorithm for HPF spectra, we provide an initial description here on its performance. As we assemble a larger HPF spectral library, we plan to further detail the performance and applicability of this algorithm on the larger high-resolution NIR M-dwarf library. The subsections below further discuss our spectral library, the algorithm and its performance on recovering the stellar parameters of the stars internal to the library using a cross-calibration scheme, our best-estimate stellar parameter values of G 9-40, and finally we discuss future work we plan to perform to further improve the performance of the code framework described here. 

\subsubsection{Spectral Library}
We assembled a spectral library of high S/N as-observed HPF spectra that compose a subsection of the library stars in the \texttt{SpecMatch-Emp} code from \cite{yee2017}, all of which have precisely characterized stellar parameters. We restrict our library to a subsection of the stars from \cite{yee2017} that have an effective temperature of $T_{\mathrm{eff}} < 4000 \unit{K}$. Our current library consists of 26 stars, where all of the spectra have a S/N>162 with a median S/N of 495 evaluated at $\sim$1.1micron. The current library covers the following parameter ranges: $3080 \unit{K} < T_{\mathrm{eff}} < 3989 \unit{K}$, $4.63<\log(g) < 5.11$ and $-0.49 < \mathrm{[Fe/H]} < 0.4$. 

To generate the stellar library, we performed a first-order deblaze correction of the HPF science and the sky calibration fibers using the dedicated HPF flat-field lamp. We then subtracted the deblaze-corrected sky contamination from the science fiber light to minimize the impact of sky-background and sky emission lines on the determination of the spectral parameters. To ensure that all of the relevant spectral features line up for the spectral comparison, we shifted the spectra first to the barycenter of the solar system using \texttt{barycorrpy} package, and subsequently shifted the spectrum to the stellar rest frame by fitting for the best-fit absolute RV value by fitting a Gaussian to the Cross-Correlation Function with a reference binary mask. The CCF binary mask is the same mask we used to calculate the absolute RV of G 9-40 further detailed in subsection \ref{ref:absrv} below. After correcting for both Doppler shifts, we then resampled the spectra using linear interpolation onto a common wavelength grid with a fixed 0.01A spacing---$\sim$4 times smaller than the smallest pixel spacing in HPF to minimize any information loss in the interpolation step---to facilitate the $\chi^2$ comparison. For the parameter retrieval, we experimented using 6 different HFP orders that were relatively clean of telluric absorption lines for the parameter retrieval, all of which returned consistent stellar parameters. This comparison is further described in Subsection \ref{sec:crossval}.

\subsubsection{Empirical SpecMatch Algorithm}
Following \cite{yee2017}, we use a $\chi^2$ metric to compare the goodness-of-fit of a target spectrum $S_{\mathrm{target}}$ to a given $S_{\mathrm{reference}}$ star spectrum in the spectral library,
\begin{equation}
\chi^2_{\mathrm{initial}} = \sum_{i=1}^{N} \left( S_{\mathrm{target},i} - S_{\mathrm{reference},i}(p,v\sin i) \right)^2,
\label{eq:chi}
\end{equation}
where $S_{\mathrm{target},i}$ is the deblazed and normalized flux of the target star and $S_{\mathrm{reference},i}(p,v \sin i)$ is the deblazed and normalized flux of the library reference star at the $i$-th element of the resampled wavelength array ($N$ in total). We rotationally broadened the reference star flux, $S_{\mathrm{reference},i}(p,v\sin i)$ by a projected rotational velocity $v \sin i$, and multiplied by a 5-th order Chebyshev polynomial denoted by the 6-element vector $p$ to remove any low-order residual variations in the deblazed spectra. For the $v \sin(i)$ rotational broadening, we adopt the broadening kernel from \cite{gray1992}, following the implementation in \cite{yee2017}. Following \cite{yee2017}, we specifically do not scale the $\chi^2$ value in Equation \ref{eq:chi} by the estimated photon-noise error bars, as the residuals of the high S/N spectra are completely dominated by systematic astrophysical and/or instrumental differences rather than our estimate of the photon noise. Using the $\chi^2$ metric from Equation \ref{eq:chi}, we loop through all of the spectra in the library to optimize for a minimum $\chi^2$ value as a function of the Chebyshev polynomial vector $p$ and $v \sin i$. We used the Nelder-Mead simplex ("\textit{Amoeba}") algorithm \citep{nelder1965} as implemented in the \texttt{scipy.optimize} package. Following \cite{yee2017}, we enforce a limited range of allowed $v \sin i$ values between $0 \unit{km\:s^{-1}}$ (no broadening) to $15 \unit{km\:s^{-1}}$ to minimize excursions to artificially high $v \sin i$ values.

Following the $\chi^2_{\mathrm{initial}}$ loop, we then select the top 5 lowest $\chi^2_{\mathrm{initial}}$ spectra to form a new composite spectrum, $S_{\mathrm{composite}} = \sum_{j=1}^5 c_j S_j(p_j,[v \sin i]_j)$, where the 5 coefficients $c = \{c_1, c_2, c_3, c_4, c_5\}$ are optimized to further minimize the $\chi^2$ of the target star and the composite spectrum according to the following $\chi^2$ metric,
\begin{equation}
\chi^2_{\mathrm{composite}} = \sum_{i=1}^{N} \left( S_{\mathrm{target},i} - \sum_{j=1}^5 c_j S_{j,i}(p_j,[v \sin i]_j) \right) ^2,
\label{eq:chi2}
\end{equation}
where $p_j$ and $[v \sin i]_j$ are the best-fit values determined from the optimization step in Equation 1 for the $j$-th best reference spectrum ($j \in \{ 1, 2, \ldots 5\}$). We fit for the first four $c$ coefficients in the optimization step and then set,
\begin{equation}
c_5 = 1-\sum_{j=1}^4 c_j,
\end{equation}
to ensure that all 5 coefficients sum up to unity. All the $c$ coefficients are further constrained to have a value between 0 and 1.

\subsubsection{Cross-validation}
\label{sec:crossval}
To check the performance of the algorithm described above on the NIR HPF spectra, we performed a cross-validation procedure consisting of removing a given spectrum from the library and comparing the recovered best-fit stellar parameter to its known library value. We then repeated this comparison for all of the stars in the library, and computed the standard deviation ($\sigma$) of the residuals between the recovered best-fit stellar parameters and the known library value for the three stellar parameters considered (i.e., computing $\sigma_{T_{\mathrm{eff}}}$, $\sigma_{\mathrm{Fe/H}}$, and $\sigma_{\log g}$). To compare the performance of different HPF orders in recovering the known parameter values, we ran this cross-validation procedure on 6 different HPF orders that are relatively clean of telluric absorption features. Table \ref{tab:crossval} summarizes the comparison between the different orders, showing the resulting $\sigma_{T_{\mathrm{eff}}}$, $\sigma_{\mathrm{Fe/H}}$, and $\sigma_{\log g}$ values. From Table \ref{tab:crossval}, we see that the different orders overall show similar scatter, with order 5 (wavelengths: [8670\AA--8750\AA]) having the lowest $\sigma_{T_{\mathrm{eff}}}$ value, and order 17 (wavelengths: [10460\AA--10570\AA]) having the lowest $\sigma_{\mathrm{Fe/H}}$ value. In addition, Table \ref{tab:crossval} shows the individual best-fit stellar parameter point estimates for G 9-40 for the same orders. The derivation of the point estimates is further discussed in Subsection \ref{sec:g940spec}.

\begin{figure}[t]
\begin{center}
\includegraphics[width=\columnwidth]{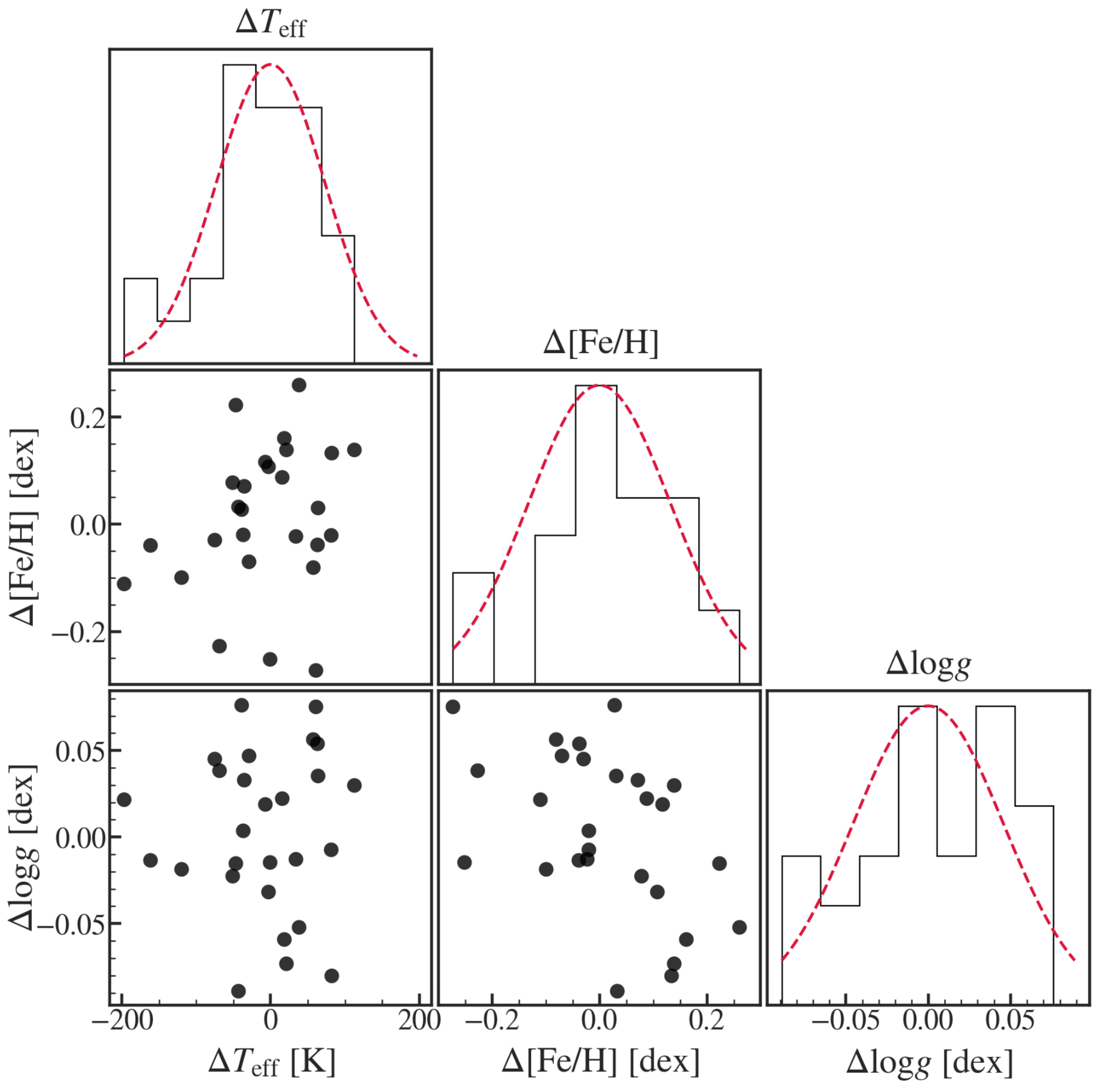}
\end{center}
\vspace{-0.3cm}
\caption{Cross-validation of the empirical spectral matching algorithm using a leave-one-out approach for HPF order 17, showing the difference (denoted by $\Delta$) between the recovered stellar-parameter value and the known value for each star in the library for $T_{\mathrm{eff}}$, [Fe/H], and $\log g$. The standard deviations of the three parameters are 73K, 0.13dex, and 0.05dex for $T_{\mathrm{eff}}$, Fe/H, and $\log(g)$, respectively. We obtain a similar level of scatter using the other HPF orders as is summarized in Table \ref{tab:crossval}. The dotted red line compares the corresponding histograms to the expected Gaussian distribution centered around zero with the same standard deviation as the observed data.}
\label{fig:crossval}
\end{figure}

Although all of the orders considered perform similarly in the cross-validation, we elect to use order 17 for our stellar parameter point estimate, as that order has the highest S/N and this order has some of the fewest telluric absorption lines of the orders considered, minimizing systematic errors from telluric absorption. Further, this order empirically shows a slightly better performance in recovering the known metallicity. Although we do not observe G 9-40 to exhibit significant chromospheric Ca Infrared (IRT) activity from inspection of the HPF spectra, we elect not to use order 5 for our final spectral parameter point estimate (although it formally shows the lowest $\sigma_{T_{\mathrm{eff}}}$ value in our cross-validation), as that order contains one of the Ca IRT lines which can be in emission for active stars and could adversely affect the $\chi^2$ comparison. As such, using the cross-validation results for order 17, we assign the $\sigma_{T_{\mathrm{eff}}}=73 \unit{K}$, $\sigma_{\mathrm{Fe/H}}=0.13 \unit{dex}$, and $\sigma_{\log g}=0.05\unit{dex}$ as our best estimate of the $1\sigma$ errors for our point estimates of $T_{\mathrm{eff}}$, [Fe/H], and $\log g$, respectively. Figure \ref{fig:crossval} graphically shows the cross-validation residuals for order 17.

\begin{deluxetable*}{ccc|ccc|ccc}
\tablecaption{Results of stellar parameter estimation for G 9-40 for the HPF orders considered (see Subsection \ref{sec:g940spec}), along with the library cross-validation residual standard deviations (see Subsection \ref{sec:crossval}) for the same orders. The cross-validation standard deviations are the standard deviation of the residuals between the recovered and known library value of all of the stars in the stellar library. We assign the cross-validation standard deviations as our error on the corresponding stellar parameter estimate. For the spectroscopic stellar-parameter point estimates, we see that all of the orders yield consistent parameters within the error bars determined from the library cross-validation. Figure \ref{fig:crossval} graphically depicts the individual library residuals for order 17, our highest S/N order, and we assign the stellar parameters derived from order 17 as our final parameters for G 9-40 as is further discussed in Subsection \ref{sec:crossval}. \label{tab:crossval}}
\tablehead{
\nocolhead{0} & \nocolhead{1} & \nocolhead{2} & \multicolumn{3}{c}{G 9-40 Best-fit Values} & \multicolumn{3}{c}{Library Cross-Validation} \\\hline
\colhead{Order}  &  \colhead{Wavelength Region}  & \colhead{S/N} & \colhead{$T_{\mathrm{eff}}$}  & \colhead{Fe/H}  & \colhead{$\log g$} & \colhead{$\sigma_{T_{\mathrm{eff}}}$}  & \colhead{$\sigma_{\mathrm{Fe/H}}$}  & \colhead{$\sigma_{\log g}$} \\
           \colhead{}       &  \colhead{[A]}     & \colhead{}   & \colhead{[K]}                 & \colhead{[dex]} & \colhead{[dex]}         & \colhead{[K]}                 & \colhead{[dex]} & \colhead{[dex]}}
\startdata
5  & [8670, 8750]   &  85 & 3426 & -0.084 & 4.896 & 64 & 0.15 & 0.05 \\
6  & [8790, 8885]   &  90 & 3420 & -0.031 & 4.860 & 83 & 0.18 & 0.06 \\
14 & [9940, 10055]  & 126 & 3366 & -0.048 & 4.878 & 100& 0.17 & 0.06 \\
15 & [10105, 10220] & 131 & 3378 & -0.063 & 4.873 & 86 & 0.17 & 0.05 \\
16 & [10280, 10395] & 134 & 3381 & -0.061 & 4.898 & 69 & 0.15 & 0.05 \\
\textbf{17} & \textbf{[10460, 10570]} & \textbf{138} & \textbf{3405} & \textbf{-0.078} & \textbf{4.909} & \textbf{73} & \textbf{0.13} & \textbf{0.05} \\
\enddata
\end{deluxetable*}

\subsubsection{G 9-40 Parameter Estimation} 
\label{sec:g940spec}
\begin{figure*}[t]
\begin{center}
\includegraphics[width=0.95\textwidth]{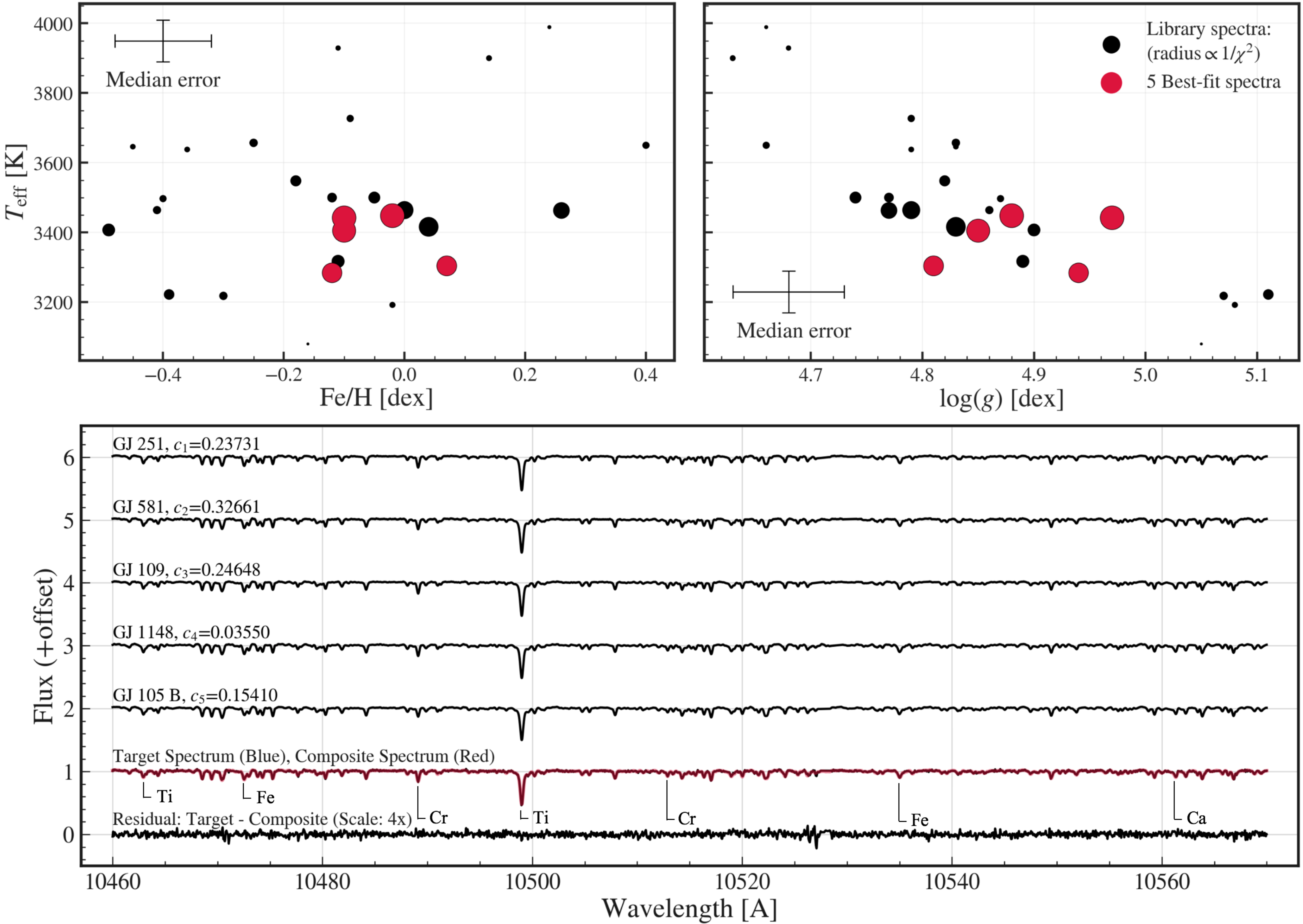}
\end{center}
\caption{\textbf{Top panels}: Best-fit library stars to G 9-40 showing the $T_{\mathrm{eff}}$ of the library stars as a function of [Fe/H] (left) and $\log g$ (right). The radius of each data point is inversely proportional to the calculated $\chi^2_\mathrm{initial}$ value from Equation \ref{eq:chi}, so larger points show a lower $\chi^2_{\mathrm{initial}}$ value and thus indicate a better fit. Highlighted in red are the 5 best matching stars: GJ 251, GJ 581, GJ 109, GJ 1148, and GJ 105 B. We use the spectra of these stars to perform a second linear-combination optimization step to interpolate between the stellar parameters of these stars. \textbf{Bottom:} Comparison of the 5 top-best-fitting library spectra, the best-fit linear combination spectrum of these 5 spectra, and the resulting residuals (target G 9-40 spectrum subtracted from the composite spectrum, scaled by a factor of 4). The optimal weights for each of the 5 library spectra are listed. A few of the atomic lines (Ti, Fe, Cr, Ca) are annotated at the bottom as retrieved from the VALD database \citep{ryabchikova2015}. Wavelengths are in vacuum wavelengths.}
\label{fig:chi2}
\end{figure*}

We ran the spectral matching algorithm described above on the highest S/N spectrum of G 9-40, which had a signal-to-noise of 148 at $\sim$1.1micron and a S/N of 137 in order 17. The top panels in Figure \ref{fig:chi2} visualize the $\chi^2_{\mathrm{initial}}$ values as calculated using Equation \ref{eq:chi} for all of the library stars in the $T_{\mathrm{eff}}$ and [Fe/H] plane (left) and the $T_{\mathrm{eff}}$ and $\log g$ plane (right). The 5 best-matching stars (GJ 251, GJ 581, GJ 109, GJ 1148 and GJ 105 B) are highlighted in red in the upper panels in Figure \ref{fig:chi2}. We then used these 5 best-matching stars for the second linear combination step optimizing the $\chi^2_{\mathrm{composite}}$ value in Equation \ref{eq:chi2}. The lower panel in Figure \ref{fig:chi2} compares the spectra of these 5 stars, along with the best-fit linearly-combined composite spectrum. The final composite spectrum and the corresponding residuals from the target star are shown at the bottom. Using the weights we determine the following stellar parameters for G 9-40 using order 17: $T_{\mathrm{eff}}=3405 \pm 73 \unit{K}$, $\mathrm{Fe/H} = -0.078 \pm 0.13$ and $\log (g) = 4.909 \pm 0.05$, where our best-estimate of the error bars is derived from our cross-validation step for order 17 in Table \ref{tab:crossval}. We note that running the algorithm independently on the other orders in Table \ref{tab:crossval}, results in consistent stellar parameter estimates within the $1\sigma$ uncertainties. 

\subsubsection{Future Work}
Although Table \ref{tab:crossval} demonstrates that the current version of our spectral matching framework with HPF spectra is performing reliably, we discuss a number of future-work topics here that could further improve its performance. This includes correcting the spectra for tellurics using dedicated telluric-removal software, which could unlock the use of other HPF orders for spectral parameter inference. Further, more thorough performance testing is needed at low S/N levels. We have explicitly left this as future work, as all of the spectra in the current library and the spectrum of G 9-40 were all of high S/N (library spectra are all higher than S/N>162, and the G 9-40 spectrum used has a S/N=148). Additionally, the algorithm can only infer parameters that are within the bounds of the stellar library. As such, we plan to extend the library to cooler M-dwarfs to enable robust spectral parameter estimation of late-type M-dwarfs using HPF spectra. Finally, as further noted above, the observed spectral residuals are completely dominated by astrophysical systematics. In the future, we plan to investigate the possibility of adapting the \texttt{Starfish} code \citep{czekala2015} to the spectral matching framework discussed here. \texttt{Starfish} implements a flexible Gaussian Process interpolator capable of deriving spectral parameters and posteriors from observed spectra given a spectral library while self-consistently accounting for correlated errors in the interpolation between library spectra.

\subsection{Spectral Energy Distribution and Isochrone Fitting}
\label{sec:sed}
\begin{figure}[t]
\begin{center}
\includegraphics[width=\columnwidth]{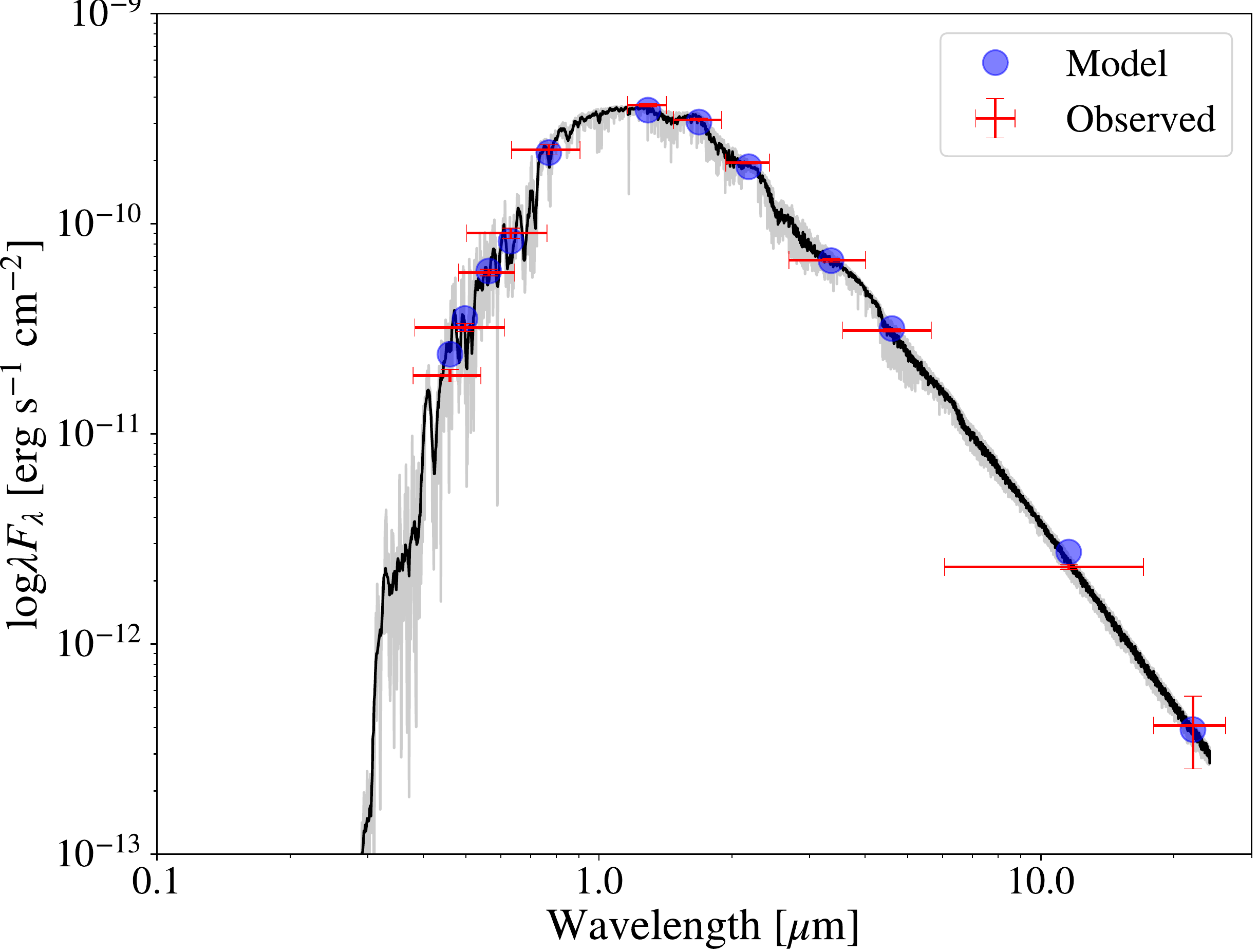}
\end{center}
\caption{The spectral energy distribution model (SED) of G 9-40. The grey line is the raw BT-NextGen model and black line is the model smoothed with a boxcar average of 10 points. The SED was fit with \texttt{EXOFASTv2} using the distance inferred from \textit{Gaia}. The error bars in wavelength reflect the bandwidth of the respective photometric filter and the error bars in flux reflect the measurement uncertainty. The blue circles are the points on the best-fitting model corresponding to the midpoint of each photometric filter. The resulting stellar parameters are summarized in Table \ref{tab:stellarparam}.}
\label{fig:sed}
\end{figure}

To estimate the mass, radius, and age of G 9-40, we use \texttt{EXOFASTv2} to perform a Spectral-Energy-Distribution (SED) and isochrone fit to the available literature photometry (see Table \ref{tab:stellarparam}), using the \textit{Gaia} parallax, and the spectroscopically determined stellar parameters discussed in the previous subsection as Gaussian priors. \texttt{EXOFASTv2} uses the BT-NextGen Model grid of theoretical spectra \citep{Allard2012} and the Mesa Isochrones and Stellar Tracks \citep[MIST,][]{dotter2016,choi2016} to fit the SED and derive model-dependent stellar parameters. To test the consistency of our spectroscopic $\log g$ measurement from our spectral-matching algorithm, we use the same $\log g = 4.909$ value as a starting point for our SED analysis but do not impose a strict prior, to minimize biases on the inferred evolutionary state, mass, and radius of the star \citep[see e.g.,][]{torres2012}. Figure \ref{fig:sed} shows the resulting SED fit. The derived model-dependent stellar parameters show good agreement with the spectroscopic values derived from the HPF spectra from the previous subsection.

\label{sec:comparison}
\cite{dressing2019} provide independent spectroscopic and photometric estimates of the stellar parameters of G 9-40 to the parameters presented here. Their spectroscopic estimates are $T_{\mathrm{eff}} = 3264_{-121}^{+137}\unit{K}$, $R_*=0.328_{-0.054}^{+0.064}\unit{R_\odot}$, $M_*=0.169_{-0.153}^{+0.142} \unit{M_\odot}$, and $\mathrm{[Fe/H]}=-0.202\pm0.084$. Their photometric parameters are $T_{\mathrm{eff}} = 3310 \pm 57\unit{K}$, $R_*=0.313 \pm 0.009 \unit{R_\odot}$, and $M_*= 0.295 \pm 0.007 \unit{M_\odot}$, where they adapt the photometric parameters as the best estimate of the stellar parameters. Their adapted parameters are in good agreement to both our spectroscopic stellar parameters and our \texttt{EXOFASTv2} model dependent parameters as summarized in Table \ref{tab:stellarparam}. 

\subsection{$v \sin i$ and Absolute RV}
\label{ref:absrv}
We measure the projected rotation velocity $v \sin i$ of the star using the high-resolution spectra from HPF, following the method in \cite{reiners2012}. This method compares the cross-correlation Function (CCF) of the target spectrum to the resulting CCFs of artificially broadened spectra of a slowly rotating template star. To calculate the CCF, we use the CCF routines from the publicly available CERES (Collection of Elemental Routines for Echelle Spectra) code base \citep{brahm2017}, which calculate the CCF by cross-correlating a given spectrum to a fixed binary mask. We elected to use Barnard's Star as the template star as it is known to be a slow rotator \citep[see e.g.,][]{ribas2018} and has a similar spectral type of M4 to G 9-40 of M2.5. To generate the binary mask for the CCF, we used a peak-finding algorithm to find single unblended lines in the Barnard's Star spectra. Blended lines have been shown to introduce substantial systematic errors in the determination of the $v \sin i$ for relatively small differences in the spectral characteristics of input stars \citep{reiners2018}. To minimize the impact of tellurics, we only run the peak finding algorithm on the 8 HPF orders cleanest of tellurics (same 8 orders as in the RV analysis in this work). Although relatively clean of tellurics, we further filtered the resulting list to produce a list of lines with minimal telluric overlap using the same telluric mask used for the RV analysis in this work (see Section \ref{sec:spectra}). To estimate the scatter in our determination of the $v \sin i$ value, we independently estimated the $v \sin i$ for each of the 8 orders separately, resulting in 8 independent point estimates of the $v \sin i$. In doing so, all independent orders resulted in a $v\sin i$ at the lower limit of our $v \sin i$ measurement precision (i.e., the width of the target star CCF was fully consistent with the width of the slow-rotator calibrator CCF), suggesting that the $v\sin i$ is below the resolution of HPF. At the resolution of HPF ($R = 55,000$, FWHM=$6\unit{km\:s^{-1}}$), we estimate that we can detect rotation velocities of about $2 \unit{km\:s^{-1}}$ or more. We thus conclude that G 9-40 has a $v \sin i < 2 \unit{km\:s^{-1}}$, from the resolution of HPF. This agrees with the slow rotational velocity observed in the \textit{K2} data, as is further discussed in the following subsection.

To measure the absolute RV of G 9-40, we fit a Gaussian profile to the CCF discussed above. Analogous to the $v\sin i$ determination, we fitted each order independently yielding 8 point estimates of the absolute RV. All 8 point estimates yielded a consistent value, and we used the standard deviation of the 8 different point estimates as our estimate of the error. This resulted in an absolute RV of $RV = 14.65 \pm 0.26 \unit{km\:s^{-1}}$, which is also summarized in Table \ref{tab:stellarparam}.

\subsection{Rotation Period}
We use the \textit{K2} photometry to estimate the rotation period of G 9-40. Looking at the photometry from \textit{K2} (Figure \ref{fig:transit}a) we see that there are small-scale ($<1\%$) modulations with a period between 25-35 days seen on top of a long-term trend after a sharp flux decrease after the first three days. We argue that both the initial sharp flux decrease and the long-term trend are likely not astrophysical (e.g., due to potential systematics including thermal settling of the spacecraft). Therefore, to estimate the rotation period, we remove the first three days and then remove a simple linear trend from the resulting photometry, yielding the light curve shown in Figure \ref{fig:transit}b. We then fit the rotation period using two methods as described below.

\begin{figure}[h!]
\begin{center}
\includegraphics[width=1.0\columnwidth]{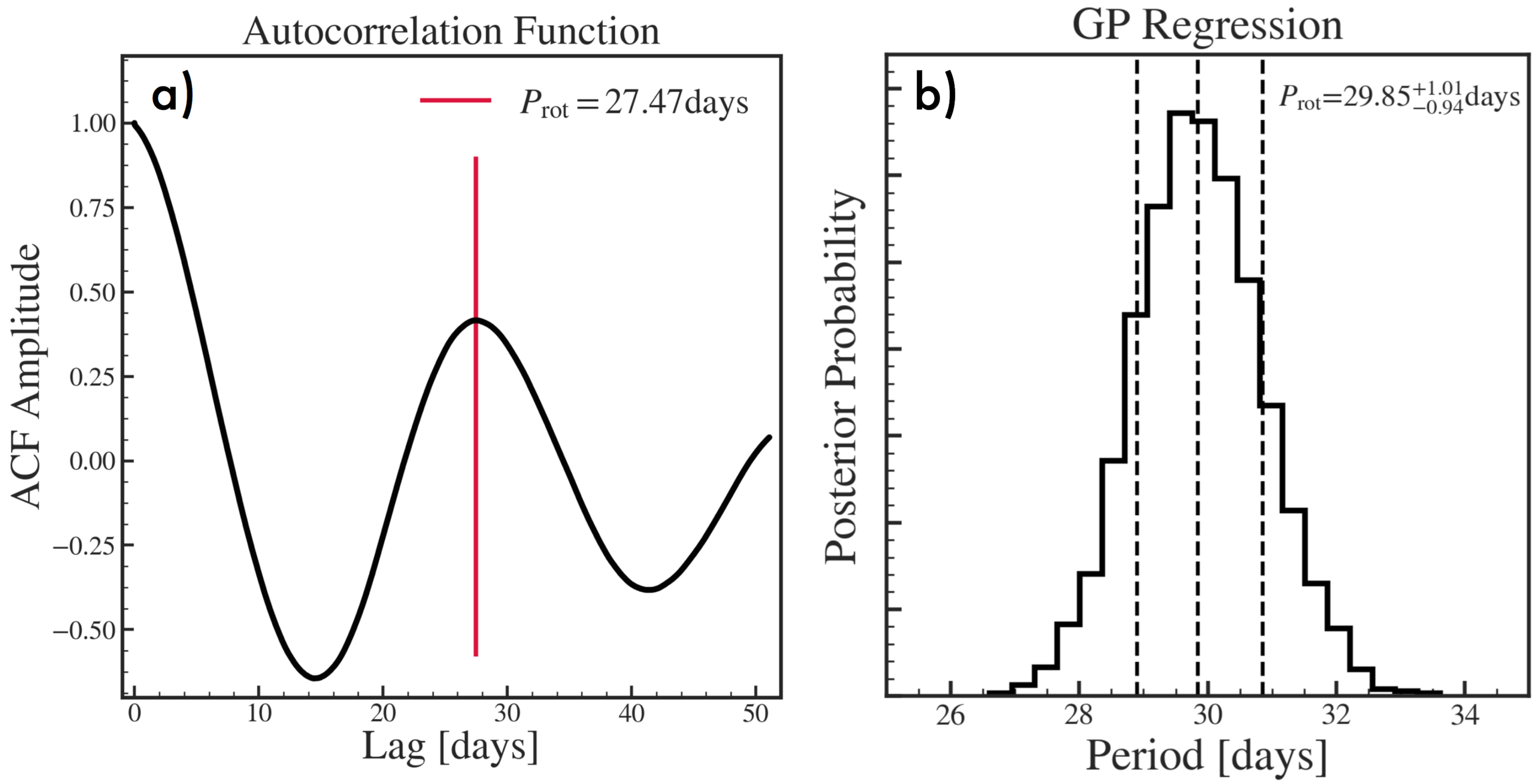}
\end{center}
\caption{a) Autocorrelation function of the \textit{K2} photometry after removing transits and removing a linear trend (photometry shown in Figure \ref{fig:transit}b). The red line shows the location of the first peak, suggesting a rotation period of $P_{\mathrm{rot,acf}}=27.47\unit{days}$. b) Rotational period posteriors from our quasi-periodic kernel GP regression of the same photometry as in a), suggesting a rotational period of $P_{\mathrm{rot,gp}}=29.85_{-0.94}^{+1.01}\unit{days}$.}
\label{fig:acf}
\end{figure}

First, we use the first peak of the autocorrelation function (ACF) as an estimate of the rotation period \citep[see e.g.,][]{mcquillan2013}. To generate the autocorrelation function, we used the \texttt{acf} function in the \texttt{statsmodels} Python package. Before calculating the ACF, we masked the datapoints during transit, removed any $>3\sigma$ outliers, and then interpolated the masked values to give a uniformly sampled light curve at the $\sim$30 minute cadence of \textit{K2}. Figure \ref{fig:acf} shows the resulting ACF for the linear-trend-removed photometry in Figure \ref{fig:transit}b along with the location of the highest peak, yielding a rotation period of $27.5 \unit{days}$. 

Second, we derived an additional estimate for the rotation period by using a Gaussian process. \cite{Angus2018} show that a quasi-periodic covariance function is effective for making probabilistic measurements of the rotation period even if the data is sparsely sampled. Exact Gaussian process modeling has the disadvantage of the runtime scaling as \(\mathcal{O}\left(N^{3}\right)\). For this purpose, we use the \texttt{juliet} analysis package \citep{Espinoza2018} which models the photometry using the formalism presented in \cite{Foreman-Mackey2017},  where a simple function is constructed that mimics the properties of the quasi-periodic covariance function, and implemented with the \texttt{celerite} Python package. The period is the only hyperparameter and there are three nuisance parameters. Using the same data which generated the ACF, we calculate a rotation period of $29.85^{+1.01}_{-0.94}$ days, which is in broad agreement with our rotational period estimate from the ACF method (within $3\sigma$). This results in a maximum stellar equatorial rotational velocity of $v_{\mathrm{eq}} \sim 500 \unit{m\:s^{-1}}$, in good agreement with the $v \sin i$ constraint in the previous subsection. From the \textit{K2}-photometry the amplitude of the modulation is $\sim$0.5\% which could be measured through extensive long-baseline ground-based photometric observations and used to confirm the $P_{\mathrm{rot}}$ value presented here. Although they do not present a rotation period estimate, \cite{pepper2008} classify G 9-40 as a long-period variable (LPV) using photometric data from the Kilodegree Extremely Little Telescope (KELT), which is consistent with the rotation period estimate presented here.

\begin{deluxetable}{llcc}
\tablecaption{Summary of stellar parameters. \label{tab:stellarparam}}
\tabletypesize{\scriptsize}
\tablehead{\colhead{~~~Parameter}                                 &  \colhead{Description}                                            & \colhead{Value}                         & \colhead{Reference}}
\startdata
\multicolumn{4}{l}{\hspace{-0.2cm} Main identifiers:}                                                                                                                                  \\
EPIC                                                              &  -                                                                & 212048748                               & Huber                  \\
LSPM                                                              &  -                                                                & J0858+2104                              & Lepine                 \\
2MASS                                                             &  -                                                                & J08585232+2104344                       & Huber                  \\
Gaia DR2                                                          &  -                                                                & 684992690384102528                      & Gaia                   \\
\multicolumn{4}{l}{\hspace{-0.2cm} Equatorial Coordinates, Proper Motion and Spectral Type:}           \\
$\alpha_{\mathrm{J2000}}$                                         &  Right Ascension (RA)                                             & 08:58:52.32                             & Gaia                   \\
$\delta_{\mathrm{J2000}}$                                         &  Declination (Dec)                                                & +21:04:34.20                            & Gaia                   \\
$\mu_{\alpha}$                                                    &  Proper motion (RA, \unit{mas\ yr^{-1}})                          & $175.512 \pm 0.103$                     & Gaia                   \\
$\mu_{\delta}$                                                    &  Proper motion (Dec, \unit{mas\ yr^{-1}})                         & $-318.469 \pm 0.067$                    & Gaia                   \\
Spectral Type                                                     &  -                                                                & M2.5                                    & Reid                   \\
\multicolumn{4}{l}{\hspace{-0.2cm} Optical and near-infrared magnitudes:}           \\
$B$                                                               &  APASS Johnson B mag                                              & $15.462 \pm 0.074$                      & APASS                  \\
$V$                                                               &  APASS Johnson V mag                                              & $13.823 \pm 0.040$                      & APASS                  \\
$g^{\prime}$                                                      &  APASS Sloan $g^{\prime}$ mag                                     & $14.626 \pm 0.047$                      & APASS                  \\
$r^{\prime}$                                                      &  APASS Sloan $r^{\prime}$ mag                                     & $13.217 \pm 0.061$                      & APASS                  \\
$i^{\prime}$                                                      &  APASS Sloan $i^{\prime}$ mag                                     & $11.994 \pm 0.061$                      & APASS                  \\ 
\textit{Kepler}-mag                                                        &  \textit{Kepler} magnitude                                                 & $12.771$                                & Huber                  \\
$J$                                                               &  2MASS $J$ mag                                                    & $10.058 \pm 0.022$                      & 2MASS                  \\
$H$                                                               &  2MASS $H$ mag                                                    & $9.433  \pm 0.023$                      & 2MASS                  \\
$K_S$                                                             &  2MASS $K_S$ mag                                                  & $9.190  \pm 0.018$                      & 2MASS                  \\
$WISE1$                                                           &  WISE1 mag                                                        & $9.032  \pm 0.023$                      & WISE                   \\
$WISE2$                                                           &  WISE2 mag                                                        & $8.885  \pm 0.020$                      & WISE                   \\
$WISE3$                                                           &  WISE3 mag                                                        & $8.774  \pm 0.029$                      & WISE                   \\
$WISE4$                                                           &  WISE4 mag                                                        & $8.597  \pm 0.411$                      & WISE                   \\
\multicolumn{4}{l}{\hspace{-0.2cm} Spectroscopic Parameters$^a$:}           \\
$T_{\mathrm{eff}}$                                                &  Effective temperature in \unit{K}                                & $3405 \pm 73$                           & This work              \\
$\mathrm{[Fe/H]}$                                                 &  Metallicity in dex                                               & $-0.078 \pm 0.13$                       & This work              \\
$\log(g)$                                                         &  Surface gravity in cgs units                                     & $4.909 \pm 0.05$                        & This work              \\
\multicolumn{4}{l}{\hspace{-0.2cm} Model-Dependent Stellar SED and Isochrone fit Parameters$^b$:}           \\
$T_{\mathrm{eff}}$                                                &  Effective temperature in \unit{K}                                & $3348 \pm 32$                           & This work              \\
$\mathrm{[Fe/H]}$                                                 &  Metallicity in dex                                               & $0.04_{-0.11}^{+0.10}$                  & This work              \\
$\log(g)$                                                         &  Surface gravity in cgs units                                     & $4.926 \pm 0.027$                       & This work              \\
$M_*$                                                             &  Mass in $M_{\odot}$                                              & $0.290 \pm 0.020$                       & This work              \\
$R_*$                                                             &  Radius in $R_{\odot}$                                            & $0.3073_{-0.0061}^{+0.0059}$            & This work              \\
$\rho_*$                                                          &  Density in $\unit{g\:cm^{-3}}$                                        & $14.11_{-0.92}^{+0.99}$                 & This work              \\
Age                                                               &  Age in Gyrs                                                      & $9.9_{-4.1}^{+2.6}$                     & This work              \\
$L_*$                                                             &  Luminosity in $L_\odot$                                          & $0.01069_{-0.00029}^{+0.00028}$           & This work              \\
$A_v$                                                             &  Visual extinction in mag                                         & $0.077_{-0.046}^{+0.034}$               & This work              \\
$d$                                                               &  Distance in pc                                                   & $27.928 \pm 0.045$                      & Gaia                   \\
$\pi$                                                             &  Parallax in mas                                                  & $35.807 \pm 0.059$                      & Gaia                   \\
\multicolumn{4}{l}{\hspace{-0.2cm} Other Stellar Parameters:}           \\
$P_{\mathrm{rot}}$                                                &  Rotational period in days                                        & $29.85^{+1.01}_{-0.94}$                 & This work              \\
$v \sin i_*$                                                      &  \multirow{2}{2.3cm}{Stellar rotational \hspace{0.3cm}\: velocity in \unit{km\ s^{-1}}}  & $<2$             & This work              \\
                                                                  &                                                                   &                                         &                        \\
$RV$                                                              &  \multirow{2}{2.3cm}{Absolute radial \hspace{0.3cm}\: velocity \unit{km\ s^{-1}} ($\gamma$)}   & $14.65 \pm 0.26$        & This work              \\
                                                                  &                                                                   &                                         &                        \\
\enddata
\tablenotetext{}{References are: Huber \citep{huber2016}, Lepine \citep{lepine2005}, Reid \citep{reid2004}, Gaia \citep{gaia2018}, APASS \citep{henden2015apass}, UCAC2 \citep{zacharias2004ucac2}, 2MASS \citep{cutri20032mass}, WISE \citep{cutri2014wise}.}
\tablenotetext{a}{Derived using our empirical spectral-matching algorithm.}
\tablenotetext{b}{{\tt EXOFASTv2} derived values using MIST isochrones with the \textit{Gaia} parallax and spectroscopic parameters in $a$) as priors.}
\end{deluxetable}

\section{Transit Analysis}
\label{sec:transits}

\subsection{Transit search in the K2 data}
\begin{figure*}[t]
\begin{center}
\includegraphics[width=\textwidth]{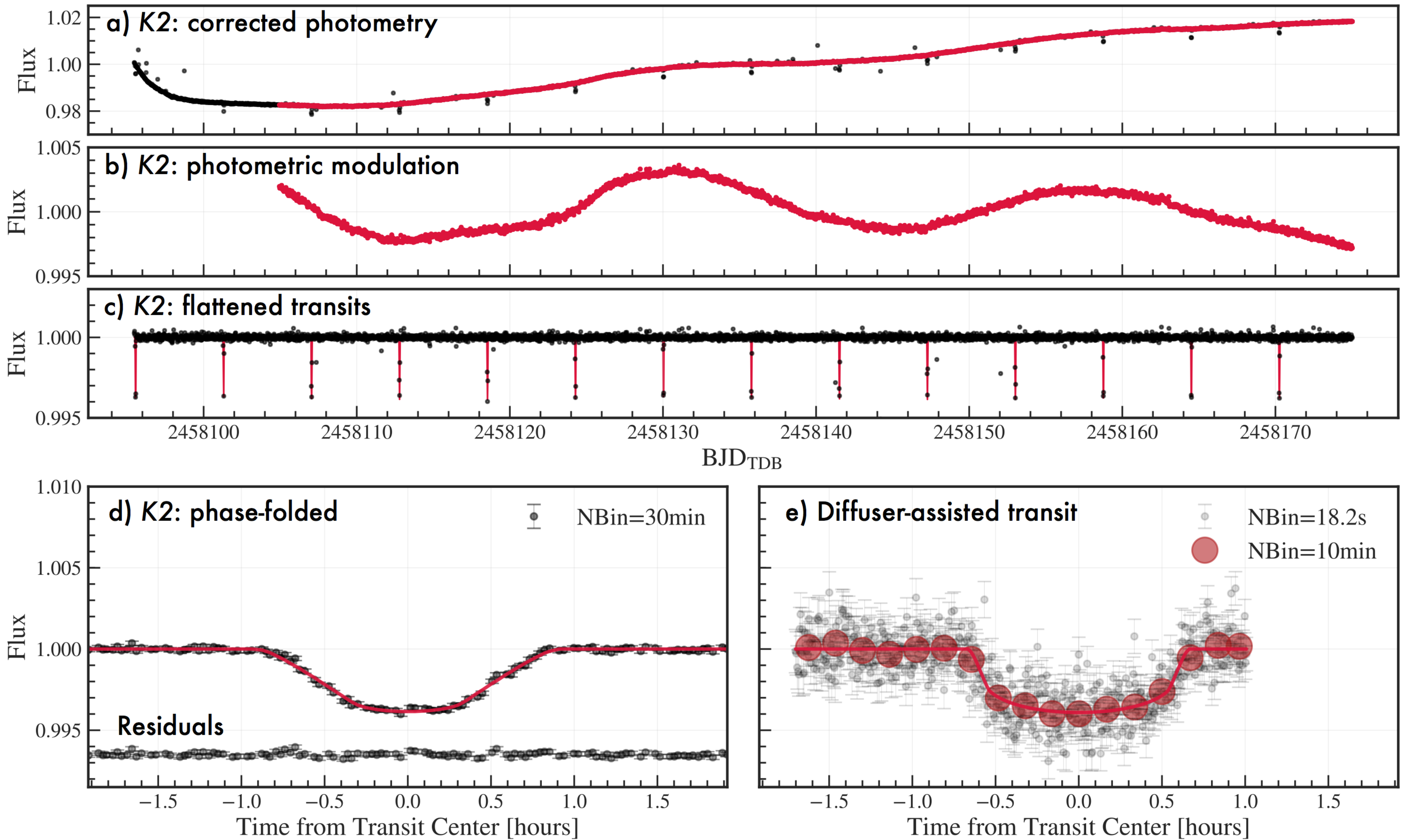}
\vspace{-0.5cm}
\end{center}
\caption{\textit{K2} and ground-based diffuser-assisted photometry of G 9-40. a) \textit{K2} photometry from the \texttt{Everest} pipeline in black after correcting the \textit{K2} data for pointing drifts and thruster events. Transits are clearly seen. The red curve is used to estimate the photometric rotation period of G 9-40 after removing a linear trend (see panel b) and any transits. b) Subset of the full \textit{K2} photometry used to estimate the photometric rotation period after removing a linear trend and the first 400 points from panel a). Further, any transits have been masked out and linearly interpolated over to generate a homogeneously sampled curve to calculate an Auto Correlation Function (ACF) to estimate the rotation period (see Figure \ref{fig:acf}). c) Flattened light curve in black (same as black curve in a) after flattening with our best-fit GP model from \texttt{Everest}. Our best-fit transit model is shown in red. d) Phase-folded \textit{K2} photometry along with the best-fit model in red. e) Ground-based diffuser-assisted photometry in the custom Semrock 857/37nm filter using the ARCTIC imager on the 3.5m ARC Telescope at APO. Black points have a cadence of 18.2s and the red points show the photometry binned to a 10min cadence.}
\label{fig:transit}
\end{figure*}

Although \cite{yu2018} provide an ephemeris for the transits in the K2 data for EPIC 212048748, we describe our independent transit search here. To search for transits, we flattened the \textit{K2} light curve of G 9-40 using the best-fit Gaussian-Process model from \texttt{Everest}. We then ran a Fortran and Python implementation\footnote{The BLS package is openly available on GitHub: \url{https://github.com/dfm/python-bls}} of the Box-Least-Square (BLS) algorithm \citep{kovacs2002} on the flattened light-curve. After finding a significant periodic signal using the BLS algorithm, we masked $2 \times T_{14}$ transit-time windows surrounding the calculated transit midpoints and then reran the BLS algorithm to look for potential further transits. We found no strong evidence for other transits in the system. The final \textit{K2} photometry from \texttt{Everest} is shown in Figure \ref{fig:transit} along with the identified transits.

\subsection{Transit fitting}
After identifying the transit locations using the BLS algorithm we ran 3 different fits: a fit of the \textit{K2} transits only, a fit of the ground-based transit only, and finally a joint fit with the \textit{K2} and ground-based transits simultaneously to put further constraints on the orbital period. We performed all transit fits in a Markov-Chain Monte-Carlo framework to obtain parameter posteriors following the methodology described in \cite{stefansson2017,stefansson2018a}. In short, for all three fits we used the \texttt{batman} Python package for the transit model which uses the transit model formalism from \cite{mandel2002}. Before starting the MCMC runs, we found the global best-fit solution using a differential-evolution global optimization package called \texttt{PyDE}\footnote{\url{https://github.com/hpparvi/PyDE}}, maximizing the log-posterior probability (sum of the log-likelihood function and log prior probabilities). We initialized 100 walkers in distributed in a uniform N-dimensional sphere (where N is the number of jump parameters used) close to the maximum likelihood solution using the \texttt{emcee} affine-invariant MCMC ensemble sampler \citep{dfm2013}. We computed 50,000 steps for each Markov Chain walker, resulting in a total of 5,000,000 computed samples. We removed the first 5,000 steps in each chain as burn-in resulting in 4,500,000 samples used for final parameter inference. The Gelman-Rubin statistic for the resulting chains were all $<3\%$ from unity which we considered well-mixed \citep[see e.g.,][]{ford2006}.

We used the following five MCMC jump parameters describing the planet transit: transit center ($T_C$), period ($\log(P)$), inclination ($\cos(i)$), radius ratio ($R_p/R_*$) and semi-major axis ($\log(a/R_*)$), along with one parameter for the out-of-transit baseline flux. We explicitly fix the eccentricity and the argument of periastron to be equal to 0. We impose broad uniform priors on all 5 jump parameters that are summarized in Table \ref{tab:priors}. For $T_C$, $\log(P)$, and $R_p/R_*$, we center the priors on the values determined by the BLS algorithm, but as we do not get constraints on the values of $\log(a/R_*)$ and $\cos(i)$ from the BLS algorithm, we imposed wide uniform priors on these parameters. We impose a Gaussian prior centered on unity for the baseline flux parameter. We used a quadratic limb-darkening law to describe the limb-darkening using the triangular-sampling parametrization described in \cite{kipping2013}. Following \cite{kipping2013}, we fully marginalize over the complete parameter space of the $q_1$ and $q_2$ limb-darkening coefficients (from 0 to 1) to minimize biases on our planet parameter values due to inaccuracies in the stellar parameters.

Table \ref{tab:priors} summarizes the priors used for all of the different fits. We report the median values along with the 16th and 84th percentile error bars derived from the posteriors for all three fits in Table \ref{tab:planetparams}. Our joint-fit median values collectively provide a good description of the transit (i.e., no obvious bimodality is seen in the resulting posteriors). We further note that our resulting best-fit transit parameters are consistent within the $1\sigma$ error bars with the parameters obtained by \cite{yu2018}. We further compare our resulting transit ephemeris to the ephemeris obtained by \cite{yu2018} in Section \ref{sec:discussion}.

\subsubsection{K2 Light Curve Analysis}
To reduce the data volume of the \textit{K2} data to be analyzed, we clipped the light curve in $2 \times T_{14}$ windows---where $T_{14}$ is the transit duration between first and fourth contact---surrounding the expected transit midpoints after flattening the light curve and removing any 4\(\sigma\) outliers) using the best-fit \texttt{Everest} GP detrending model. Using the \texttt{exposure\_time} keyword in the \texttt{batman} package, we oversampled and binned the model to reflect the 30 minute \textit{Kepler} cadence as suggested by \citep{kipping2010}. For the transit modeling, we fixed the error bar per photometric observation to the standard deviation of all of the points outside the transit but within the $2 \times T_{14}$ transit window, resulting in a fixed error estimate of $\sigma_{\mathrm{30min,K2}}=130\unit{ppm}$. In addition to this, in our MCMC fits we also experimented fitting for the optimal \textit{K2} error bar by including the error bar as a free parameter in the \textit{K2} fit, which resulted in a consistent error estimate. As both values agreed, we elected to keep the \textit{K2} error bar estimate fixed, to reduce the number of free parameters in the fit. The MCMC jump parameters and associated priors for this fit are summarized in Table \ref{tab:priors}.

\subsubsection{Ground-based Light Curve Analysis}
For the ground-based analysis we estimate the photometric uncertainty, including scintillation, following the discussion surrounding Equation 3 in \cite{stefansson2018a}, yielding error bars that account for photon, dark, readout, background, digitization and scintillation noise. For the scintillation error calculation, we assumed that the single reference star used was fully uncorrelated to the scintillation error from the target star. The jump parameters, along with the associated priors for this fit, are summarized in Table \ref{tab:priors}. In addition to the transit jump parameters, in this MCMC fit we also simultaneously detrend with a line to account for a low-amplitude linear slope observed in the photometry. 

\begin{figure}
\begin{center}
\includegraphics[width=0.99\columnwidth]{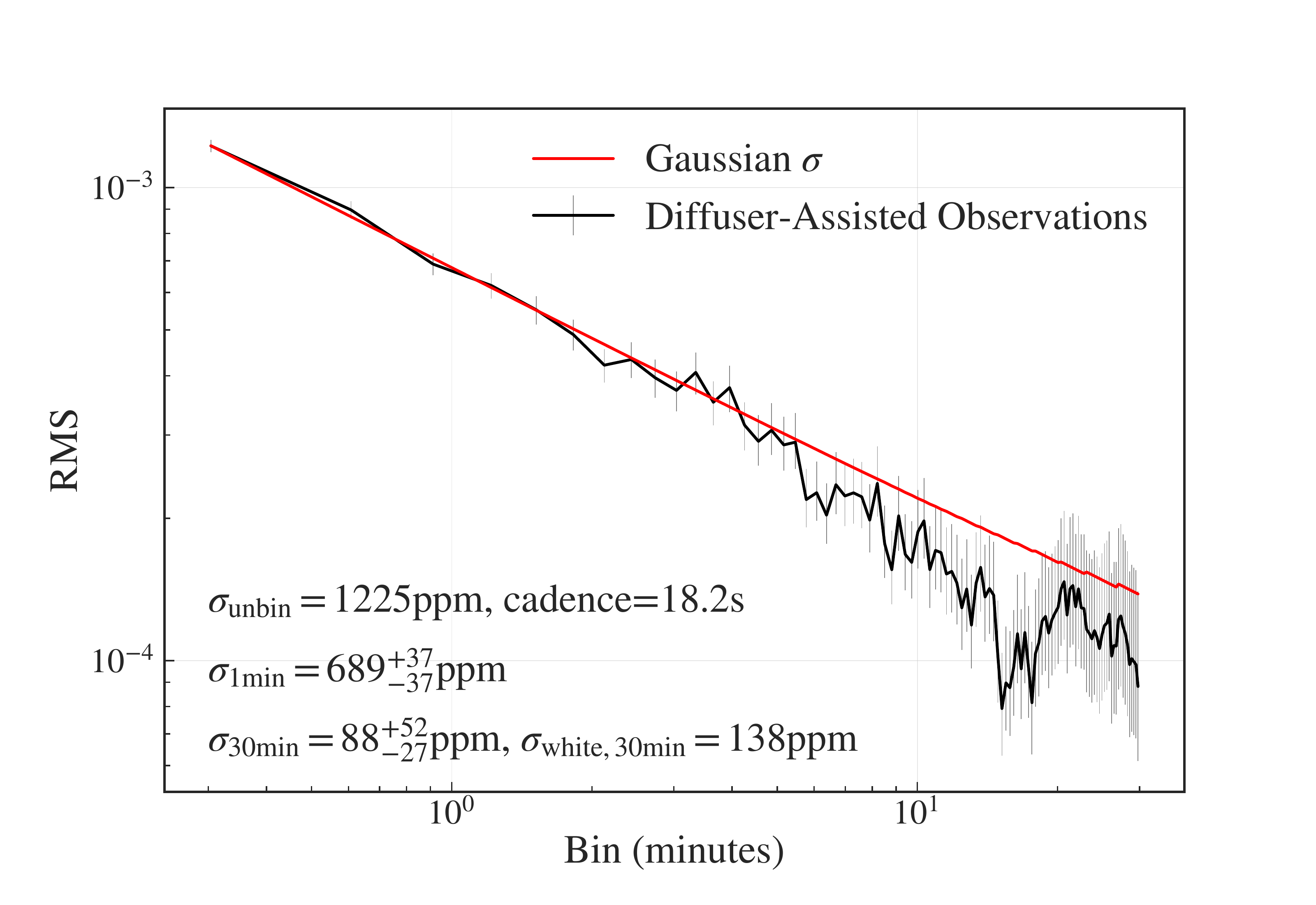}
\vspace{-0.5cm}
\end{center}
\caption{The root-mean-square (RMS) value of the residuals of the ground-based diffuser-assisted photometry as a function of bin size in minutes (black curve) compared to the Gaussian expected bin-down behavior shown in red. We see that the data largely bin down as Gaussian white-noise, with some downward excursions below the expected Gaussian behavior seen at the largest bin sizes. We conservatively say that we reach a precision of 138ppm in 30min.}
\label{fig:rms}
\end{figure}

In Figure \ref{fig:rms}, we show the photometric precision (or the root-mean-square (RMS) of the best-fit residuals) as a function of bin size using the \texttt{MC3} package from \cite{cubillos2017}, which estimates error bars on the RMS values assuming they follow an inverse gamma distribution. Additionally, shown in red in Figure \ref{fig:rms} is the expected bin-down behavior assuming Gaussian white-noise. From Figure \ref{fig:rms}, we see that in our ground based observations, we achieve a photometric precision of $\sigma_{\mathrm{unbin}}=1225\unit{ppm}$ unbinned (18.2 second cadence), which bins down to  $\sigma_{\mathrm{1min}}=689\pm37\unit{ppm}$ and $\sigma_{\mathrm{30min}}=88_{-27}^{+52}\unit{ppm}$, in 1 minute and 30 minute bins, respectively. For comparison, we note that the Gaussian expected precision in 30 minutes is 138ppm. We observe that the RMS values largely follow the expected white-noise behavior, although we note that slight excursions below the Gaussian expected values are observed at the largest bin sizes, which still are largely consistent with the Gaussian expected precision within the reported error bars. Similar and larger departures below the Gaussian expected precision have been reported by a number of groups in the literature \citep[e.g.,][]{blecic2013,stefansson2017}, and \cite{cubillos2017} show that these excursions are not statistically significant after taking into account the increasingly skewed inverse gamma distribution of the RMS values at the largest bin sizes. Following our previous work \citep{stefansson2017,stefansson2018a}, we argue that excursions much below the Gaussian expected precision is likely an overestimate of the actual precision achieved, and we conservatively say that we achieve 138ppm precision in 30 minute bins for these transit observations.

\subsubsection{Joint K2 and Ground-based Light Curve Analysis}
To further constrain the planet parameters, in addition to the individual \textit{K2} and ground-based only fits discussed above, we performed a joint fit of both the \textit{K2} and the ground-based observations. For this fit, we imposed uniform priors on the period and the transit center, and we assume that the planet follows a strictly periodic orbit with no transit timing variations\footnote{We see no obvious signs of transit timing variations (TTVs) in the phase folded K2 transits, and the transit center of our diffuser-assisted observations is consistent with our $K2$-only linear ephemeris (see Subsection \ref{sec:ephemeris}).}. Following \cite{stefansson2018a}, to account for variations in the transit depth due to a potential planetary atmosphere, we allowed the radius ratio $R_p/R_*$ to vary separately for the \textit{K2} and ground-based data, while assuming common values for the $\log(P)$, $T_C$, $\log{a/R_*}$, and $\cos (i)$ parameters. We include independent parameters for the transit baselines, and perform a simultaneous linear detrend for the ground-based observations as discussed in the previous subsection. Similarly, for the \textit{K2}-only fit, we fix the \textit{K2} error bar at 130ppm for each individual photometric point from \textit{K2}. The priors we used for this fit are summarized in Table \ref{tab:priors}. Our best-fit joint models are overlaid in red in Figure \ref{fig:transit} in panels c, d, and e. 

\begin{deluxetable*}{llccc}
\tablecaption{Summary of priors used for the three MCMC transit fits performed. $\mathcal{N}(m,\sigma)$ denotes a normal prior with mean $m$, and standard deviation $\sigma$; $\mathcal{U}(a,b)$ denotes a uniform prior with a start value $a$ and end value $b$. The eccentricity was assumed to be 0 for all fits. For all fits we uniformly sampled quadratic limb-darkening parameters $q_1$ and $q_2$ from 0 to 1 using the formalism from \cite{kipping2013}. Priors on $T_\mathrm{eff}$, $R_*$ are adapted from Table \ref{tab:stellarparam}, but no prior is placed on the stellar density $\rho$. \label{tab:priors}}
\tablehead{\colhead{Parameter}&  \colhead{Description}                   & \colhead{\textit{K2}-only}              & \colhead{Ground-only}                 & \colhead{Joint}    } 
\startdata	
\hline
$\log(P)$ (days)              &  Orbital period                          & $\mathcal{U}(0.7592,0.7594)$            & $\mathcal{N}(0.75936,0.00002)$        & $\mathcal{U}(0.7592,0.7594)$                 \\ 
$T_C$                         &  Transit Midpoint $(\mathrm{BJD_{TDB}})$ & $\mathcal{U}(2458095.54,2458095.56)$    & $\mathcal{U}(2458497.76,2458497.80)$  & $\mathcal{U}(2458497.76,2458497.80)$         \\ 
$(R_p/R_*)_{\mathrm{K2}}$     &  Radius ratio                            & $\mathcal{U}(0,0.1)$                    & --                          		   & $\mathcal{U}(0,0.1)$                         \\ 
$(R_p/R_*)_{\mathrm{ground}}$ &  Radius ratio                            & --                            	       & $\mathcal{U}(0,0.1)$                  & $\mathcal{U}(0,0.1)$                         \\ 
$\cos(i)$                     &  Transit inclination                     & $\mathcal{U}(0,0.2)$         	       & $\mathcal{U}(0,0.2)$                  & $\mathcal{U}(0,0.2)$                         \\ 
$\log(a/R_*)$                 &  Normalized orbital radius               & $\mathcal{U}(0.9,2.0)$       	       & $\mathcal{U}(0.9,2.0)$                & $\mathcal{U}(0.9,2.0)$                       \\ 
fraw$_{\mathrm{K2}}$          &  Transit baseline for \textit{K2} data   & $\mathcal{U}(0.9,1.1)$                  & --                                    & $\mathcal{N}(1.00002,0.0002)$                \\ 
fraw$_{\mathrm{Ground}}$      &  Transit baseline for ground-based data  & --                           	       & $\mathcal{U}(0.9,1.1)$                & $\mathcal{N}(1.001,0.001)$                   \\ 
$D_{\mathrm{Line}}$           &  Ground detrend parameter: line          & --           			               & $\mathcal{U}(-0.1,0.1)$               & $\mathcal{N}(0.00104,0.00001)$               \\ 
\enddata
\end{deluxetable*}

\begin{deluxetable*}{llcccc}
\tablecaption{Median values and 68\% confidence intervals for the transit fit parameters for our \textit{K2}-only, ground-based-only and joint \textit{K2} and ground-based MCMC analyses. For our joint fit, we fit for the transit depth ($\delta$) and the planet radius ratio ($R_p/R_*$) separately for \textit{K2} and the diffuser-assisted observations, resulting in slightly different values derived for the planet radii and transit depths in the two different bands. These values are denoted by (\textit{K2}) and (ground), respectively. We fixed the eccentricity and argument of periastron to 0 for all fits. \label{tab:planetparams}}
\vspace{-0.2cm}
\tablehead{                  \colhead{~~~Parameter}&                      \colhead{Description}  &               \colhead{\textit{K2}}    & \colhead{Ground}                       &  \colhead{Joint fit (adopted)}}
\startdata
                    $T_{C}$ $(\mathrm{BJD_{TDB}})$ &                            Transit Midpoint &  $2458095.55737_{-0.00030}^{+0.00030}$ &  $2458497.77751_{-0.00036}^{+0.00036}$ &  $2458497.77747_{-0.00033}^{+0.00032}$ \\
                                        $P$ (days) &                              Orbital period &       $5.745951_{-0.00004}^{+0.00004}$ &        $5.74596_{-0.00026}^{+0.00026}$ &     $5.746007_{-0.000006}^{+0.000006}$ \\
                         $(R_p/R_*)_{\mathrm{K2}}$ &                  Radius ratio (\textit{K2}) &            $0.059_{-0.0026}^{+0.0035}$ &                                      - &           $0.0605_{-0.0028}^{+0.0026}$ \\
                     $(R_p/R_*)_{\mathrm{ground}}$ &                       Radius ratio (ground) &                                      - &           $0.0635_{-0.0047}^{+0.0031}$ &           $0.0605_{-0.0033}^{+0.0032}$ \\
                   $R_{p, \mathrm{K2}} (R_\oplus)$ &                 Planet radius (\textit{K2}) &               $1.981_{-0.095}^{+0.12}$ &                                      - &              $2.025_{-0.097}^{+0.096}$ \\
               $R_{p, \mathrm{ground}} (R_\oplus)$ &                      Planet radius (ground) &                                      - &                 $2.13_{-0.16}^{+0.12}$ &                 $2.03_{-0.11}^{+0.11}$ \\
                         $\delta_{p, \mathrm{K2}}$ &                 Transit depth (\textit{K2}) &        $0.00348_{-0.00030}^{+0.00043}$ &                                      - &        $0.00365_{-0.00033}^{+0.00032}$ \\
                     $\delta_{p, \mathrm{ground}}$ &                      Transit depth (ground) &                                      - &        $0.00403_{-0.00058}^{+0.00041}$ &        $0.00366_{-0.00039}^{+0.00039}$ \\
                                           $a/R_*$ &                   Normalized orbital radius &                   $29.8_{-7.6}^{+5.8}$ &                   $22.4_{-2.9}^{+6.4}$ &                   $27.0_{-3.7}^{+5.4}$ \\
                                          $a$ (AU) &    Semi-major axis (from $a/R_*$ and $R_*$) &            $0.0425_{-0.011}^{+0.0082}$ &            $0.032_{-0.0042}^{+0.0092}$ &           $0.0385_{-0.0053}^{+0.0078}$ \\
 $\rho_{\mathrm{*,transit}}$ ($\mathrm{g/cm^{3}}$) &                             Density of star &                  $15.1_{-8.9}^{+11.0}$ &                    $6.4_{-2.2}^{+7.2}$ &                   $11.2_{-4.0}^{+8.3}$ \\
                                  $i$ $(^{\circ})$ &                         Transit inclination &                $88.88_{-0.97}^{+0.79}$ &                $87.98_{-0.49}^{+0.85}$ &                $88.57_{-0.47}^{+0.63}$ \\
                                               $b$ &                            Impact parameter &                 $0.58_{-0.38}^{+0.23}$ &               $0.788_{-0.20}^{+0.064}$ &               $0.672_{-0.22}^{+0.100}$ \\
                                               $e$ &                                Eccentricity &                          0.0 (adopted) &                          0.0 (adopted) &                          0.0 (adopted) \\
                             $\omega$ ($^{\circ}$) &                      Argument of periastron &                          0.0 (adopted) &                          0.0 (adopted) &                          0.0 (adopted) \\
                             $T_{\mathrm{eq}}$ (K) &        Equilibrium temp. (assuming $a=0.3$) &                $304.0_{-26.0}^{+48.0}$ &                $350.0_{-42.0}^{+26.0}$ &                $319.0_{-28.0}^{+25.0}$ \\
                             $T_{\mathrm{eq}}$ (K) &        Equilibrium temp. (assuming $a=0.0$) &                $434.0_{-37.0}^{+69.0}$ &                $500.0_{-59.0}^{+37.0}$ &                $456.0_{-40.0}^{+35.0}$ \\
                                $S$ ($S_{\oplus}$) &                             Insolation Flux &                    $5.9_{-1.8}^{+4.7}$ &                   $10.4_{-4.1}^{+3.4}$ &                    $7.2_{-2.2}^{+2.5}$ \\
                                   $T_{14}$ (days) &                            Transit duration &           $0.0546_{-0.0018}^{+0.0026}$ &           $0.0583_{-0.0025}^{+0.0026}$ &           $0.0557_{-0.0017}^{+0.0019}$ \\
                                     $\tau$ (days) &                     Ingress/egress duration &           $0.0044_{-0.0015}^{+0.0044}$ &           $0.0086_{-0.0040}^{+0.0035}$ &           $0.0056_{-0.0019}^{+0.0023}$ \\
                    $T_{S}$ $(\mathrm{BJD_{TDB}})$ &                   Time of secondary eclipse &  $2458098.43034_{-0.00028}^{+0.00028}$ &  $2458500.65049_{-0.00038}^{+0.00039}$ &  $2458500.65047_{-0.00033}^{+0.00032}$ \\
\enddata
\end{deluxetable*}

\section{Statistical Validation and False Positive Analysis}
\label{sec:vespa}
To estimate the probability that the transits observed by \textit{K2} were due to astrophysical false positives, we used the \texttt{VESPA} \citep{morton2012vespa,morton2015vespa} code. The \texttt{VESPA} algorithm statistically validates a planet by simulating and determining the likelihood of a range of astrophysical false positive scenarios that could replicate the observed light curve. \texttt{VESPA} generates a population of 20,000 systems for each false positive scenario, which includes background eclipsing binaries (BEBs), eclipsing binaries (EBs), and hierarchical eclipsing binaries (HEBs), to calculate the likelihoods. As input to \texttt{VESPA} we used the (i) 2MASS \(JHK\), SDSS \(g'r'i'\), and \textit{Kepler} magnitudes, (ii) \textit{Gaia} parallax, (iii) host star temperature, surface gravity, and metallicity derived from our HPF observations, and (iv) the maximum visual extinction from estimates of Galactic dust extinction \citep{Green2019}. These values are summarized in Table \ref{tab:stellarparam}.

Two additional constraints needed for the \texttt{VESPA} statistical analysis are the maximum separation for a background eclipsing object and the maximum depth of the occultation. We adopted the maximum radius from the \texttt{EVEREST} photometric aperture (16\arcsec) as the maximum allowed separation between the G 9-40 and the true source of the transit event. To obtain the maximum occultation depth, we followed the procedure used by \cite{dressing2017}. We phase-folded the \textit{K2} photometry to the period of G 9-40b and fit a light curve model to the data. The occultation was forced to have the same duration as the transit but was not constrained to be circular such that the center was allowed to float between phases \(0.3-0.7\). We record the depth at various points in this region of the light curve and adopt the maximum value as the maximum occultation depth. We performed the analysis with \texttt{VESPA} on the \textit{K2} light curves with the \(R_{p}/R_{*}\) ratio in the \textit{Kepler} bandpass set to the 16th, 50th, and 84th percentiles of the posterior distribution from this work. The results for various false positive scenarios considered as shown in Table \ref{tab:vespa}. This analysis yielded a False Positive Possibility (FPP) of  $\mathrm{FPP} < 1 \times 10^{-6}$ from which we conclude statistically validates G 9-40b as a planet.

Although not included directly in the VESPA analysis, our HPF radial velocities, which are further discussed in Section \ref{sec:rvs}, further bolster the planetary interpretation as no significant RV signal is detected. We use these RVs to provide an upper limit on the mass of G 9-40b, as is further discussed in Section \ref{sec:rvs}.

\begin{deluxetable*}{ccccc}
\tablecaption{The \texttt{VESPA} False Positive Probabilities (FPP) for various false positive scenarios (EB: Eclipsing Binary, HEB: Hierarchical Eclipsing Binary, and BEB: Background Eclipsing Binary) calculated using the 16th, 50th, and 84th percentiles of the \(R_{p}/R_{*}\) posterior probability distribution from our joint \textit{K2} and ground-based transit fit described in Section \ref{sec:transits}. All other input values are identical in each run. The values reported are the mean and standard deviation of a bootstrap of 10 samples for each individual run. \label{tab:vespa}}
\tablehead{
\colhead{\(R_{p}/R_{*}\) Percentile} &  \colhead{EB FPP}              & \colhead{HEB FPP}             & \colhead{BEB FPP}            & \colhead{Total FPP}}
\startdata
16                                   & \((29.8\pm5.9)\times10^{-9}\)  & \((11.2\pm7.6)\times10^{-8}\) & \(<1\times10^9\)                   & \((14.2\pm7.4)\times10^{-8}\) \\
50                                   & \((27.1\pm4.9)\times10^{-9}\)  & \((4.3\pm2.6)\times10^{-8}\)  & \((1.0\pm1.4)\times10^{-8}\) & \((8.1\pm2.8)\times10^{-8}\)  \\
84                                   & \((36.5\pm7.5)\times10^{-9}\)  & \((12.7\pm6.1)\times10^{-8}\) & \((1.8\pm2.0)\times10^{-8}\) & \((18.2\pm6.4)\times10^{-8}\) \\
\enddata
\end{deluxetable*}

\section{Discussion}
\label{sec:discussion}

\subsection{Updated Ephemeris}
\label{sec:ephemeris}
\begin{figure}[h!]
\begin{center}
\includegraphics[width=0.95\columnwidth]{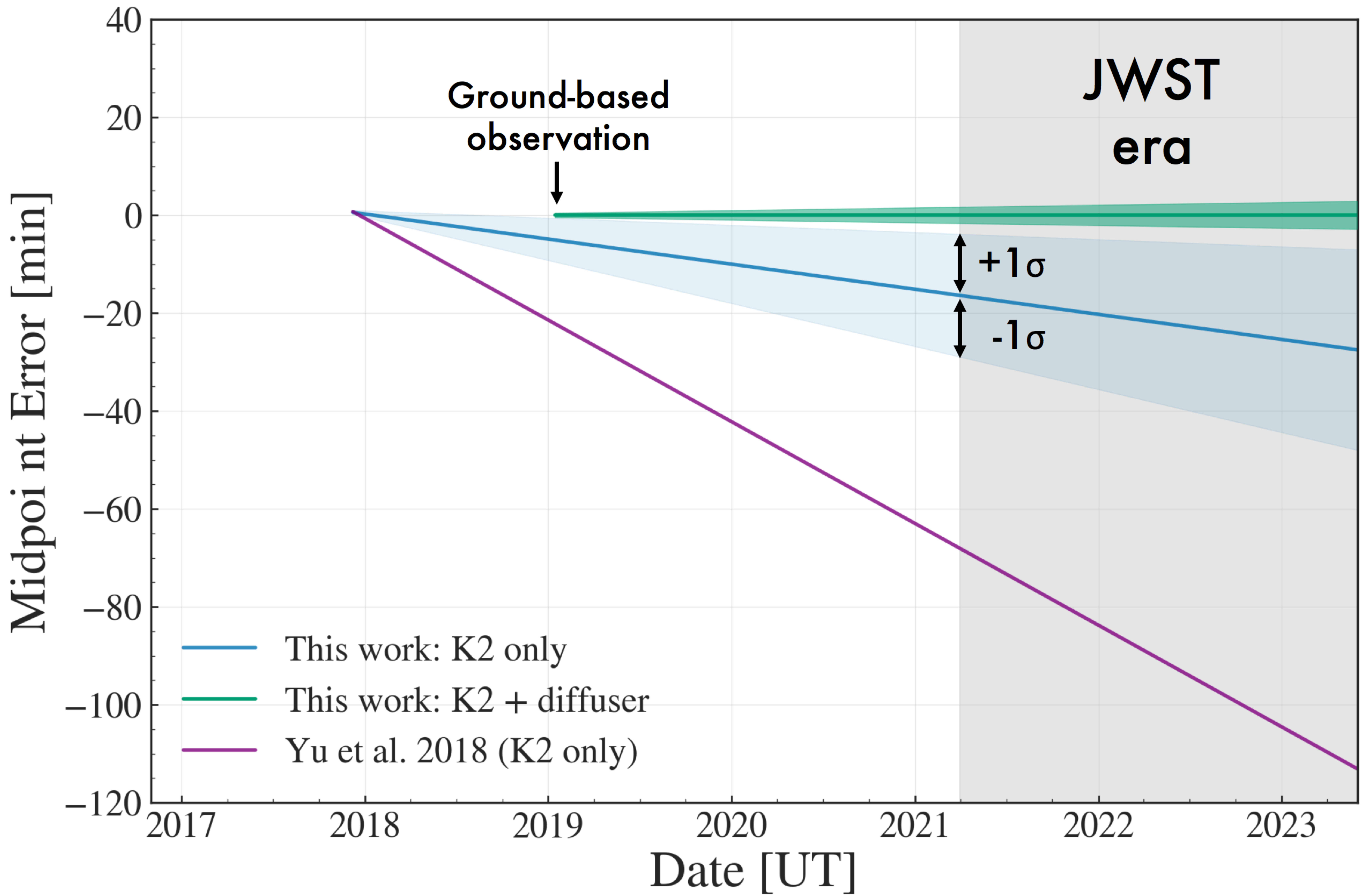}
\vspace{-0.2cm}
\end{center}
\caption{Evolution of the G 9-40b transit ephemeris uncertainties into the JWST era. The blue and green curves show the transit ephemeris derived from our \textit{K2}-only fit, and our joint \textit{K2} and diffuser-assisted transit fits, respectively. Additionally shown are the associated $1\sigma$ error bands in the shaded areas. We show that our joint \textit{K2} and diffuser-assisted fit reduces the expected uncertainty in the ephemerides in the JWST era from $\sim$13min down to $\sim$1.7min, allowing for efficient scheduling of future characterization observations during transit. Additionally shown is the transit ephemeris from \cite{yu2018} in purple (no uncertainty estimate is available), which is discrepant to the ephemeris reported here.}
\label{fig:ephemeris}
\end{figure}

To enable better future scheduling of transit observations, we fit our ground-based diffuser transit and the \textit{K2} transits together to provide an updated ephemeris of G 9-40b. Figure \ref{fig:ephemeris} compares errors on the transit center as a function of time into the beginning of the JWST era. From our \textit{K2}-only fit (blue curve), we see that the expected errors are $\sim$13min at the start of the JWST era in 2021. However, from our updated join-fit ephemeris (green curve) we reduce this error by a factor of 8 down to $\sim$1.7min. We further see that our ground-based transit was observed at the upper edge of our $1\sigma$ error bar estimate from our \textit{K2}-only fit.

In Figure \ref{fig:ephemeris} we additionally compare our ephemerides to the ephemerides presented in \cite{yu2018} calculated from \textit{K2} Campaign 16 data only (purple curve). We see that there is a discrepancy between the ephemerides derived in this work and the ephemerides presented in \cite{yu2018}. As mentioned above, our ground-based transit was observed at the upper edge of our $1\sigma$ error bar estimate from our \textit{K2}-only fit, whereas our ground-based transit was observed $\sim$23min later than expected from the \cite{yu2018} ephemeris (no uncertainty estimate is reported in \cite{yu2018} on the ephemeris). We attribute the discrepancy to the different photometric reduction and detrending methods, and we prefer the ephemeris reported here for the planning of future transit observations, given the consistency of the \textit{K2}-only fits and our ground-based transit observations.

\subsection{Stellar Density}
We used our joint transit model to get an independent estimate of the stellar density, using the equation from \citep{seager2003},
\begin{equation}
\rho_{*,\mathrm{transit}} = \frac{3 \pi}{G P^2}{\left(\frac{a}{R_*}\right)}^3,
\end{equation}
where G is the gravitational constant, and the eccentricity is assumed to be 0. From Table \ref{tab:stellarparam}, we see that our stellar density derived from our stellar radius and mass ($\rho_* = 14.11_{-0.92}^{+0.99} \unit{g\:cm^{-3}}$) is consistent with the stellar density derived from our transit observations ($\rho_{*,\mathrm{transit}} = 11.2_{-4.0}^{+8.3} \unit{g\:cm^{-3}}$).

\subsection{Predicted Most-likely Mass Estimate}
\begin{figure}[t]
\begin{center}
\includegraphics[width=\columnwidth]{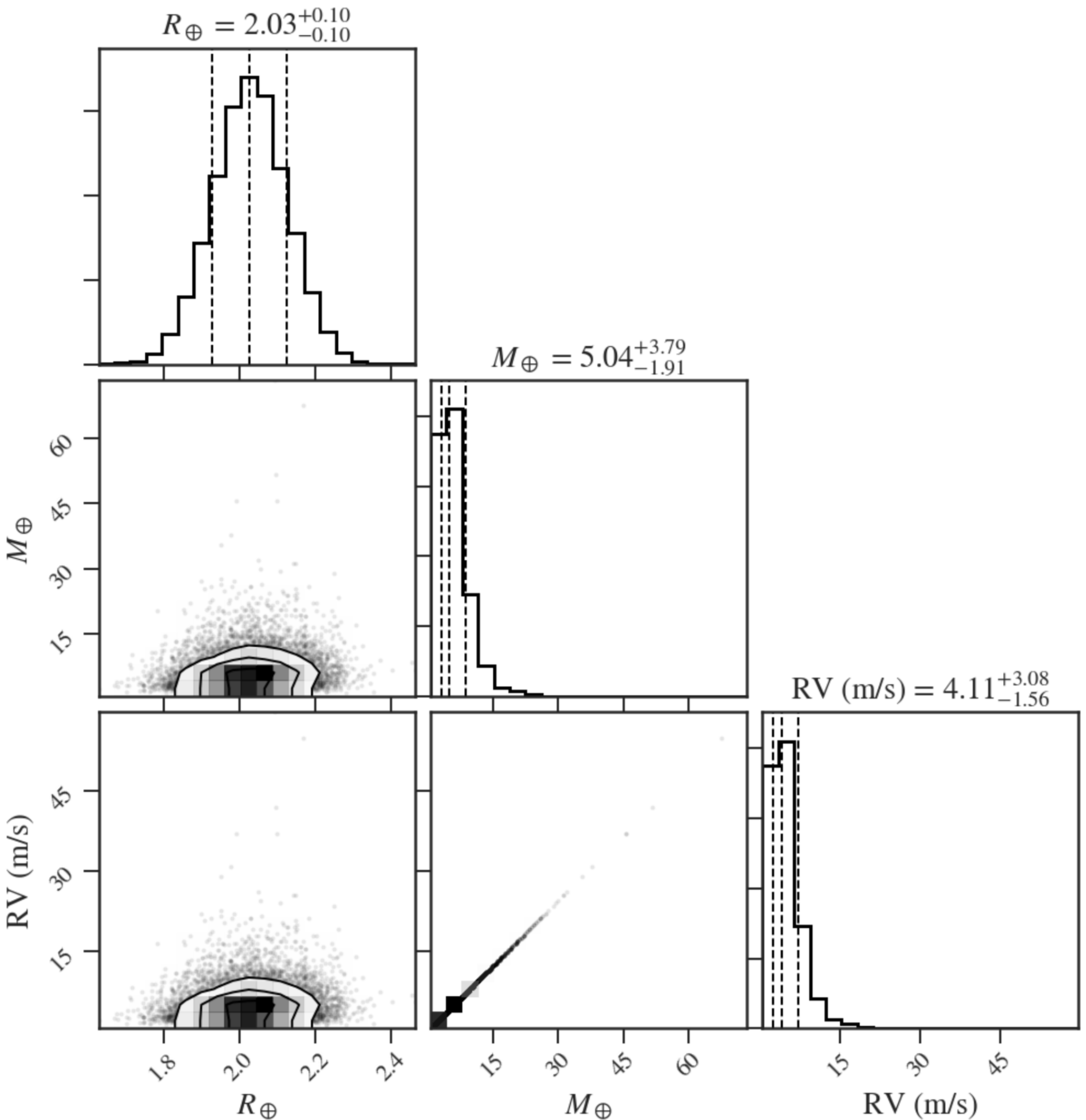}
\end{center}
\vspace{-0.2cm}
\caption{Probabilistic RV semi-amplitudes and masses calculated from the measured radii using \texttt{Forecaster} \citep{chen2017}. The expected mass and semi-amplitude are $M=5.04^{+3.79}_{-1.91} M_\oplus$, and $K=4.11^{+4.62}_{-2.41} \unit{m\:s^{-1}}$, respectively. This value agrees well with the mass estimate of $M=3.94^{+11.5}_{-3.3} M_\oplus$ we calculate using the \texttt{MRExo} M-dwarf exoplanet Mass-Radius relationship.}
\label{fig:rvmasses}
\end{figure}

We predict the most-likely mass of G 9-40b using two different methods. First, we use the \texttt{Forecaster} Python package \citep{chen2017}, which adopts a broken-powerlaw model to model the exoplanet Mass-Radius (MR) relation and is capable of predicting the masses of exoplanets given their radii. Using \texttt{Forecaster}, we obtain an expected mass of $M=5.04^{+3.79}_{-1.91} M_\oplus$ which yields an expected RV semi-amplitude of $K= 4.11^{+3.08}_{-1.56} \unit{m\:s^{-1}}$ assuming a circular orbit. Figure \ref{fig:rvmasses} shows the expected mass and RV-semiamplitude posteriors as calculated using \texttt{Forecaster}. Further, using these posteriors, \texttt{Forecaster} is capable of classifying the planet as Terran, Neptunian, Jovian or stellar. From the posteriors in Figure \ref{fig:rvmasses}, \texttt{Forecaster} classifies G 9-40b as Neptunian with $99.7\%$ confidence.

Second, we compare our predicted mass from Forecaster to the value from the \texttt{MRExo} toolkit. As is discussed in \cite{kanodia2019}, \texttt{MRExo} offers two different MR relations for forecasting: a \textit{Kepler}-planet-sample MR relation, and an M-dwarf-planet MR relation. For this fit, we used the M-dwarf-planet MR relation given the M2.5 spectral type of the host star, yielding a predicted mass of $M=3.94^{+11.5}_{-3.3} M_\oplus$. Although both values are within the formal error bars of each other, we note that the median value from \texttt{MRExo} is smaller than the $M=5.04^{+3.79}_{-1.91} M_\oplus$ value we obtain from \texttt{Forecaster}. Further, the \texttt{MRExo} value has a larger spread than the value from \texttt{Forecaster}. This larger spread is noted in \cite{kanodia2019} as due to the non-parametric fitting technique and the low number of M-dwarf-only planets in the M-dwarf MR relation used by \texttt{MRExo}, resulting in a less-precise estimate. 

Assuming a median semi-amplitude of $4\unit{m\:s^{-1}}$, this shows that G 9-40b can be followed up with a number of high-precision RV instruments already online or under construction \citep{fischer2016,wright2017}. As noted by \cite{kanodia2019}, there are rising statistical trends that suggest a difference between the exoplanet MR relation between M-dwarfs and exoplanets orbiting earlier type stars. However, the comparison of the two populations is still limited by the low-number of M-dwarf planets with both precise radii and masses. With a measurement of its mass, G 9-40b will yield further direct insights into those statistical trends.

\begin{figure*}[t]
\begin{center}
\includegraphics[width=\textwidth]{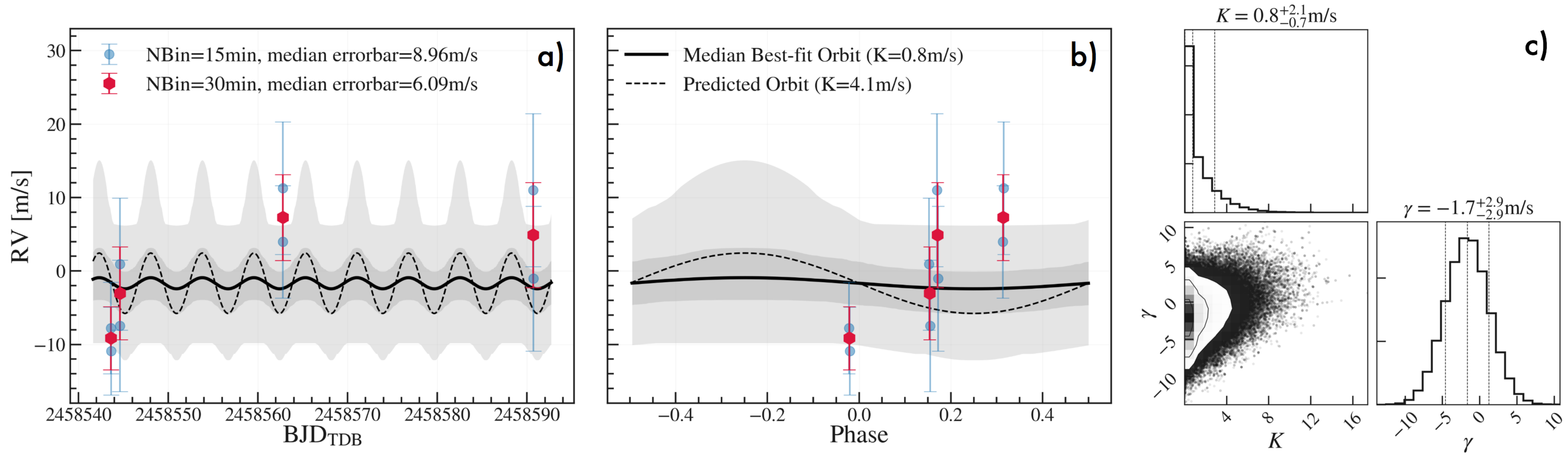}
\end{center}
\vspace{-0.2cm}
\caption{Upper limit on the mass of G 9-40b using RVs from HPF assuming a circular orbit. a) Radial velocities of G 9-40 from HPF shown as a function of time. Blue points show unbinned RVs (15 min exposures) and red points show RVs binned per HET track (30min exposure equivalent). The solid black curve shows the expected orbit using the ephemeris from Table \ref{tab:planetparams} and using our median predicted mass of $5.04 M_\oplus$ with a semi-amplitude of $K=4.11 \unit{m\:s^{-1}}$, estimated using the \texttt{Forecaster} package. The grey shaded regions show the $1\sigma$ (dark grey) and $3\sigma$ (light grey) envelopes surrounding our median best-fit MCMC model (shown in the solid black curve) of the RVs. b) Same as in a), but showing the RVs phased to the best-fit transit ephemeris from Table \ref{tab:planetparams}. c) Posterior correlations between the semi-amplitude $K$ and zero-point offset $\gamma$ from our best-fit MCMC fit to the RVs in a) and b). Using the $K$ posteriors from the assumed circular orbit, we conclude that $K<9.0 \unit{m\:s^{-1}}$ with 99.7\% ($3\sigma$) confidence, which corresponds to a mass constraint of $M<11.7 M_\oplus$ at 99.7\% confidence.}
\label{fig:rvs}
\end{figure*}

\subsection{Upper bound on Mass Estimate using the HPF RVs}
\label{sec:rvs}
To further constrain possible false positive binary scenarios and to provide an upper bound of the mass of G 9-40b, Figure \ref{fig:rvs}a shows the RVs of G 9-40 as a function of time as observed by HPF (Table \ref{tab:rvs} lists the RVs in tabular format). The blue points are individual 945s exposures ($\sigma_{\mathrm{unbinned}}=6.49 \unit{m\:s^{-1}}$), whereas the red points show weighted-average RVs per individual HET track ($\sigma_{\mathrm{binned}}=5.32 \unit{m\:s^{-1}}$; effective exposure time of 2x 945s = 31.5minutes). The associated error bars as derived from the SERVAL pipeline are $9.13 \unit{m\:s^{-1}}$ and $6.23 \unit{m\:s^{-1}}$ for the unbinned and binned points, respectively. Figure \ref{fig:rvs}b shows the phase-folded RVs using our best-fit ephemeris from our joint-fit in Table \ref{tab:planetparams}, showing that our RVs are predominately located in phases between $0$ and $0.4$. Further, for comparison, also shown in Figure \ref{fig:rvs}a and b, is the predicted RV-curve using our best prediction of the mass of G 9-40b given its radius using the \texttt{Forecaster} package ($M=5.04^{+3.79}_{-1.91} M_\oplus$, and $K=4.11^{+3.08}_{-1.56} \unit{m\:s^{-1}}$) assuming a circular orbit. 

To estimate the upper bound on the mass of G 9-40b, we modeled the RVs in Figure \ref{fig:rvs} using the \texttt{radvel} Python package. For our RV model, we fixed the orbital period and transit center to the joint-fit transit ephemeris in Table \ref{tab:rvs}, and given the few number of RV points we additionally fixed the orbital eccentricity and argument of periastron to 0. This resulted in a two-parameter RV model using only the orbital semi-amplitude ($K$) and a zero-point offset ($\gamma$) as the MCMC jump-parameters. For the modeling, we placed minimal uninformative priors on the two parameters, only constraining $K$ to be positive. To sample the parameter space, we initialized 100 MCMC walkers in the vicinity of the expected best-fit solution using \texttt{radvel} MCMC interface. We removed the first marginally well-mixed iterations as burn-in, and ran the MCMC chains until the Gelman-Rubin statistic was within $0.1\%$ of unity, which we consider indicative of well-mixed chains. Figure \ref{fig:rvs}c shows the resulting posteriors of both $K$ and $\gamma$. From our fit, we see that the best-fit semi-amplitude for the orbital semi-amplitude is $K=0.8_{-0.7}^{+2.1}\unit{m\:s^{-1}}$ (black solid curve in Figure \ref{fig:rvs}a and b), which is consistent with 0 within the $\sim$$1\sigma$ lower error estimate. Although the median-estimated semi-amplitude from this analysis is lower than our predicted semi-amplitude from both \texttt{Forecaster} and \texttt{MRExo} as discussed above, it is within the 2$\sigma$ error estimates. We attribute the low semi-amplitude we estimate here to a combination of the few number of RV points along with their sparse phase-coverage of the full RV phase curve, and argue that more RVs are needed to further test the robustness of this measurement. The semi-amplitude posteriors reported here could be biased by uncorrected astrophysical systematics (e.g., stellar activity), instrumental systematics (e.g., persistence in the HPF H2RG detector, see \cite{metcalf2019} or \cite{ninan2019} for a discussion of systematics in NIR RVs from H2RG detectors), or model-mismatch systematics (e.g., eccentric system and/or another unknown planet could be present in the system). With these caveats in mind, we use the resulting posteriors on $K$ to say that, given the current dataset and assuming a circular orbit, the semi-amplitude of G 9-40b is $K<9.0 \unit{m\:s^{-1}}$ at 99.7\% ($3\sigma$) confidence. Using our $M=0.290 \pm 0.020 M_\odot$ mass-estimate of the host star from Table \ref{tab:stellarparam}, this translates to an upper limit mass constraint of $M<11.7 M_\oplus$ at 99.7\% confidence assuming a circular orbit. If instead we adopt the same priors as discussed above, but let the eccentricity and argument of periastron float (sampled as $\sqrt{e} \cos \omega$ and $\sqrt{e} \sin \omega$ in \texttt{radvel}), our best-fit semi-amplitude increases to $K=21_{-12}^{+19}\unit{m\:s^{-1}}$, due to the fit favoring a high eccentricity solution of $e\sim0.67$ due to the sparse phase sampling of the RVs. Even in the unlikely scenario of such a high eccentricity orbit, this translates to an upper mass limit of $\sim$$0.5 M_{\mathrm{J}}$ at 99.7\% confidence, showing that G 9-40 is a planetary mass object. We prefer the $e=0$ mass constraint, as we expect G 9-40 to be fully circularized. We estimated a circularization timescale of $\sim$150Myr for G 9-40b, following the methodology in \cite{bodenheimer2001}, assuming a tidal quality factor of $Q\sim10^3$ for a mini-Neptune mass planet \citep[extrapolating between $Q\sim10^2-10^6$ for Earth and Jupiter in the Solar system;][]{goldreich1966}, which is substantially smaller than our SED estimated age of $9.9_{-4.1}^{+2.6} \unit{Gyr}$ as summarized in Table \ref{tab:stellarparam}.

\begin{deluxetable}{lcccc}
\tablecaption{Radial velocities of G 9-40 from HPF. All exposures are 945s in length. The derivation of the RV Drift Error---the contribution of the RV drift of HPF to the total estimated RV error---is described in Appendix A. The total RV error ($\sigma_{\mathrm{RV}}$) is composed of the errorbar given by the SERVAL RV pipeline added in quadrature to our RV Drift Error estimate. The dataset is available in a machine-readable table in the on-line journal.\label{tab:rvs}}
\tablehead{\colhead{$\unit{BJD_{TDB}}$}  &  \colhead{RV}                   & \colhead{$\sigma_{\unit{RV}}$} & \colhead{RV Drift Error}  & \colhead{S/N} \\
           \colhead{}                    &  \colhead{$(\unit{m\:s^{-1}})$} & \colhead{$(\unit{m\:s^{-1}})$}  & \colhead{$(\unit{m\:s^{-1}})$}  &\colhead{$@1100 \unit{nm}$} }
\startdata
 2458543.6173586124 &   -7.78 &     6.23 & 0.02 &  140.03 \\
 2458543.6284200638 &  -10.90 &     5.96 & 0.02 &  148.49 \\
  2458544.622642529 &    0.94 &     8.99 & 0.31 &  107.54 \\
 2458544.6348675573 &   -7.47 &     8.94 & 0.18 &  106.49 \\
  2458562.782749085 &    3.98 &     7.65 & 0.20 &  131.85 \\
  2458562.794474252 &   11.27 &     9.04 & 0.22 &  113.95 \\
 2458590.6907325243 &   11.00 &    10.43 & 0.10 &  103.96 \\
 2458590.7019076305 &   -1.04 &     9.85 & 0.11 &  108.93 \\
\enddata
\end{deluxetable}

\subsection{Potential for Future Study}
\begin{figure*}[t]
\begin{center}
\includegraphics[width=0.75\textwidth]{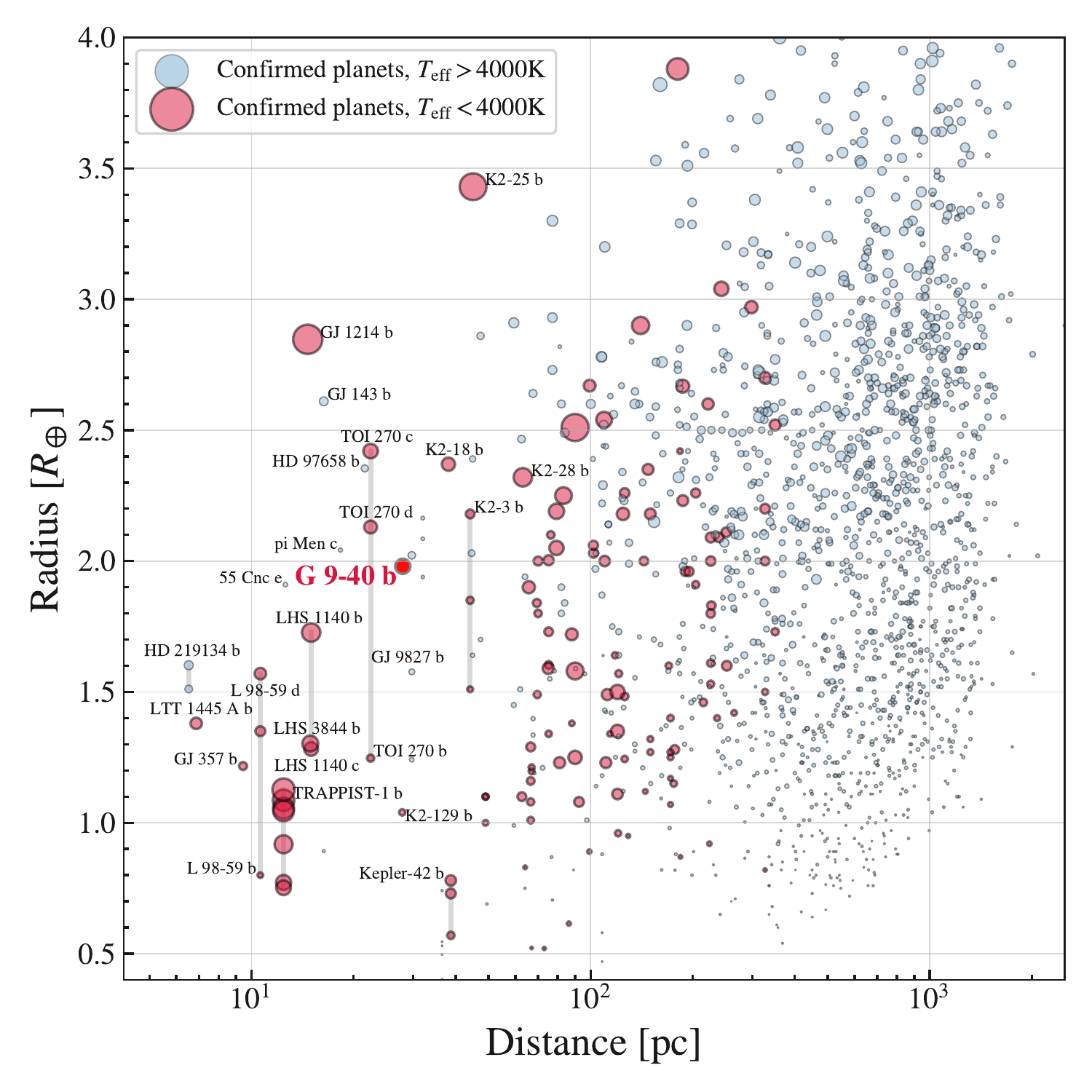}
\end{center}
\vspace{-0.5cm}
\caption{Planet radius as a function of distance for currently known planets from the NASA Exoplanet Archive (June 2019, in addition to the newly discovered M3 system L 98-59 from \cite{cloutier2019}) with radii between $0.5 R_\oplus < R < 4.0 R_\oplus$. G 9-40b (highlighted in bold red font) is the second closest planet discovered by \textit{K2} to date. M-dwarf planets are shown in red, while planets around earlier-type stars are shown in blue. The size of the points is proportional to the transit depth of the planet. Names of a few select nearby systems are shown. Grey lines connect planets in the same system for a few select M-dwarf systems. Plot inspired by Figure 4 in \cite{vanderspek2019}.}
\label{fig:distance}
\end{figure*}

Figure \ref{fig:distance} compares G 9-40b to other known planets from the NASA Exoplanet Archive \citep{akeson2013} in the exoplanet radius, and exoplanet distance from the Solar System plane. At a distance of $d=27.928 \pm 0.045 \unit{pc}$, G 9-40b is the second closest transiting planet discovered by \textit{K2} to date, behind the \textit{K2}-129 system from \cite{dressing2017} which has a Gaia DR2 distance of $27.796_{-0.074}^{+0.075} \unit{pc}$. 

Being a nearby system, G 9-40 is bright at both visible and NIR wavelengths---for example, G 9-40 is brighter in I band (I=10.9) than GJ 1214 (I=11.1), and only slightly fainter than GJ 1214 (J=9.8) in J band with a J magnitude of J=10.1. Due to its relatively large transit depth and the brightness of its host star at NIR wavelengths, G 9-40b is amenable for transmission spectroscopic observations that could provide insight into the planet's bulk composition and formation history through measuring the elemental composition of its atmosphere and overall metal enrichment. To compare the applicability of performing precision transmission spectroscopic observations of G 9-40 during transit, we calculate the Transmission Spectroscopy Metric (TSM) from \cite{kempton2018} for G 9-40b and compare it to the corresponding TSM metric for other known exoplanets. Specifically, the TSM metric from \cite{kempton2018} is defined as
\begin{equation}
\mathrm{TSM} = (\mathrm{Scale\:Factor}) \times \frac{R_p^3 T_{\mathrm{eq}}}{M_p R_*^2} \times 10^{-m_J/5},
\end{equation}
where $R_p$ is the radius of the planet in Earth radii, $M_p$ is the mass of the planet in Earth masses, $R_*$ is the radius of the host star in solar radii, $T_{\mathrm{eq}}$ is the equilibrium temperature in Kelvin of the planet calculated for zero albedo and full heat redistribution, and $m_J$ is the magnitude of the host star in $J$ band. \cite{kempton2018} define the scale factor for different radius bins where

\begin{align*}
\mathrm{(Scale\:Factor)} &= 0.19 \qquad \mathrm{for} \qquad R<1.5R_\oplus, \\
\mathrm{(Scale\:Factor)} &= 1.26 \qquad \mathrm{for} \qquad 1.5R_\oplus<R<2.75R_\oplus, \\
\mathrm{(Scale\:Factor)} &= 1.28 \qquad \mathrm{for} \qquad 2.75R_\oplus<R<4.0R_\oplus.
\end{align*}

Further, \cite{kempton2018} define the TSM metric specifically for JWST/NIRISS as NIRSS gives more transmission spectroscopy information per unit observing time compared to the other JWST instruments for a wide range of planets and host stars \citep{batalhaline2017,howe2017}. For the planets with an unknown mass, we forecast the mass using the prescription suggested in \cite{kempton2018}: using the mass-radius relationship of \texttt{Forecaster} as implemented by \cite{louie2018}.

\begin{figure*}
\begin{center}
\includegraphics[width=0.8\textwidth]{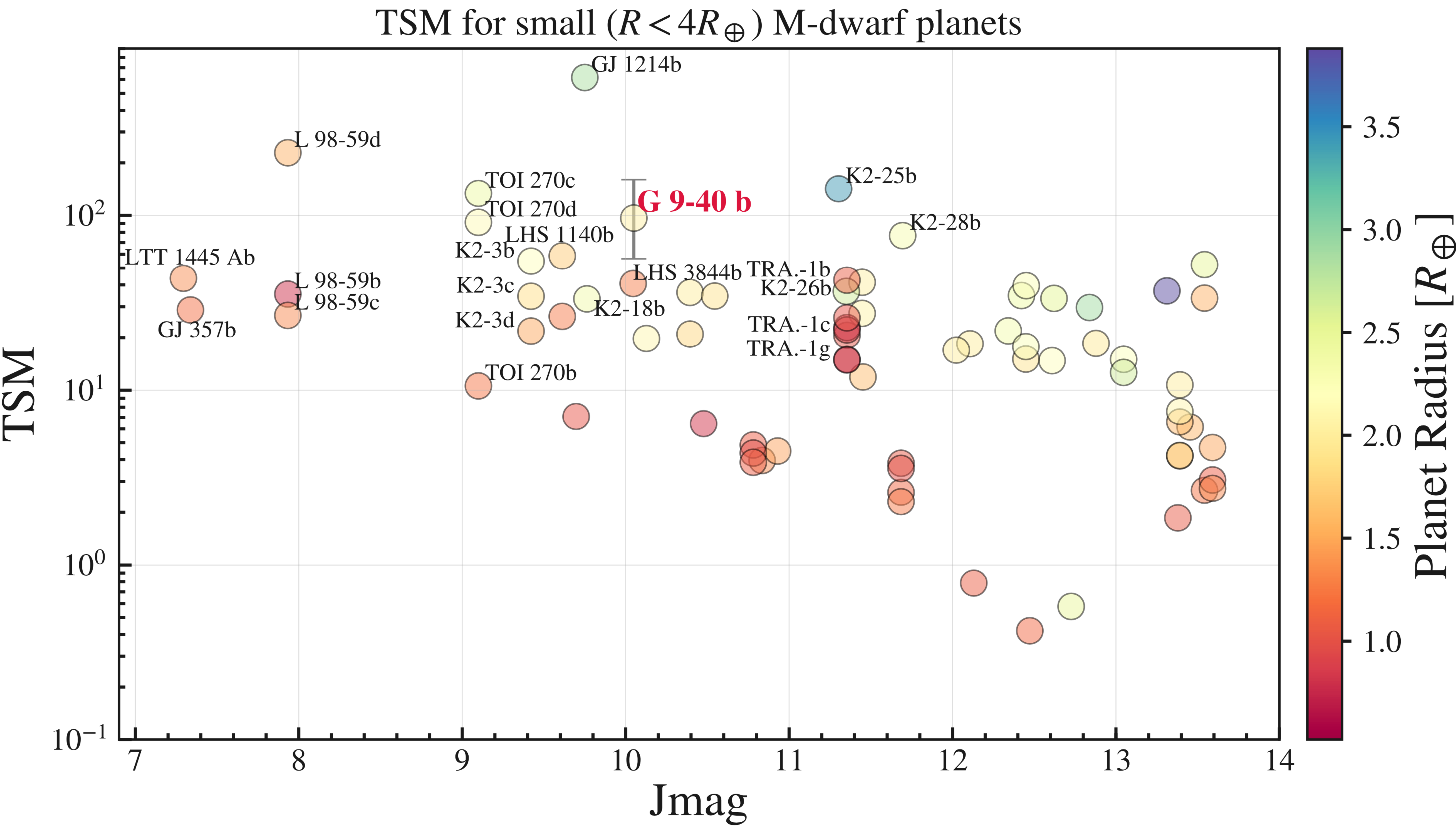}
\end{center}
\vspace{-0.5cm}
\caption{The Transmission Spectroscopy Metric (TSM) from \cite{kempton2018} as a function of J magnitude for currently known small ($<4 R_\oplus$) M-dwarf planets, with a subset of the planetary systems highlighted. G 9-40b is currently among the top M-dwarf planets for transmission spectroscopic observations with this metric in this radius bin with a TSM metric of $\mathrm{TSM}=96_{-41}^{+64}$ (assuming the predicted \texttt{Forecaster} mass posterior), exceeding the $TSM>90$ threshold for high-quality atmospheric targets suggested in \cite{kempton2018}. The TRAPPIST-1 planets are denoted by 'TRA-1'.}
\label{fig:transmission}
\end{figure*}

Using this metric---with G 9-40b falling in the "Small sub-Neptune" ($1.5 R_\oplus < R_p < 2.75 R_\oplus$) planet size bin---G 9-40b has a high TSM metric of $\mathrm{TSM}=96_{-41}^{+64}$, assuming a predicted mass of $M=5.04 M_\oplus$ using \texttt{Forecaster}, and exceeds the $\mathrm{TSM}>90$ threshold for high-quality priority atmospheric targets suggested by \cite{kempton2018}. Figure \ref{fig:transmission} compares the TSM of G 9-40b to the TSM for other currently known planets orbiting M-dwarfs ($T_{\mathrm{eff}}<4000 \unit{K}$\footnote{The known planet sample was obtained from the NASA Exoplanet Archive with the addition of the newly discovered M3 planetary system L 98-59 from \cite{cloutier2019}.}) with radii less than $4.0 R_\oplus$ as a function of the host star $J$ magnitude. This plot shows that G 9-40b is currently among the top best small (<$4.0 R_\oplus$) M-dwarf planets for transmission spectroscopic observations, after GJ 1214b \citep{charbonneau2009} ($\mathrm{TSM}\sim 615$), the newly discovered mini-Neptune ($R=1.57 R_\oplus$) L 98-59 d orbiting a nearby M3 dwarf \citep{cloutier2019} ($\mathrm{TSM}\sim 230$), the Neptune-sized ($R=3.5 R_\oplus$) M-dwarf planet K2-25b in the Hyades \citep{mann2016} ($\mathrm{TSM}\sim 140$), and the newly discovered sub-Neptune ($R= 2.42 R_\oplus$) TOI 270c \citep{gunther2019} ($\mathrm{TSM} \sim 130$).

We note that even though G 9-40b has a high TSM metric, the atmosphere of G 9-40b has the possibility of having damped atmospheric features due to its low equilibrium temperature of $T_{\mathrm{eq}}=319 \unit{K}$ ($a=0.3$). Past work has shown that hotter planets tend to have larger spectral features while cooler planets tend to have less prominent features due to hazes \citep{crossfield2017}. This has been attributed to the fact that that at temperatures below $\sim$1000K methane is abundant which can easily photolyze to produce carbon hazes which mute atmospheric features. Even if the atmosphere of G 9-40 is observed to be featureless, as noted by \cite{kempton2018}, the study of a large sample of sub-Neptune exoplanet atmospheres is especially important because these planets do not exist in our Solar System, and therefore no well-studied benchmark objects exist. Further, \cite{kempton2018} argue that a large sample of sub-Neptune planets is needed for precise atmospheric characterization, to adequately study the high degree of diversity expected for their atmospheric compositions.

\section{Summary}
We validate the discovery of a sub-Neptune-sized $R\sim2R_\oplus$ planet orbiting the nearby ($d=27.9\unit{pc}$) high proper motion M2.5-dwarf G 9-40, discovered using photometric data from the \textit{K2} mission. G 9-40 is the second closest transiting planetary system discovered by the \textit{K2} mission to date. We validate the planetary origin of the \textit{K2} transits using ground-based AO imaging using the ShaneAO system on the Lick 3m Telescope, precision diffuser-assisted narrow-band (37nm wide filter centered at 857nm) transit photometry using the ARCTIC imager at the ARC 3.5m Telescope at APO, and using precision NIR radial velocities from the HPF spectrograph on the 10m HET. 

With its large transit depth of $\sim$3500ppm and the brightness of its host star, G 9-40b is a promising target for transmission spectroscopic observations. Using the Transmission Spectroscopy Metric of \cite{kempton2018}, we show that G 9-40b is currently among the best small ($R<4R_\oplus$) M-dwarf planet for transmission spectroscopy with JWST. Using the HPF radial velocities, we place an upper bound of $<11.7 M_\oplus$ at $3\sigma$ on its mass. We predict the mass of G 9-40b using both \texttt{Forecaster} and the \texttt{MRExo} mass-radius forecasting toolkits, resulting in mass estimates of $M=5.04^{+3.79}_{-1.91} M_\oplus$ and $M=3.94^{+11.5}_{-3.3} M_\oplus$, respectively. The expected mass from \texttt{Forecaster} translates to a semi-amplitude of $K\sim 4 \unit{m\:s^{-1}}$, making G 9-40b measurable with current precision RV facilities in the red-optical and the NIR. We urge further ground-based RV follow-up to measure the mass of G 9-40b. A mass measurement of G 9-40b will provide valuable insights into the exoplanet mass-radius relationship of M-dwarf planets, along with enabling future precise transmission spectroscopic analysis.

\acknowledgments
We thank the anonymous referee for a thoughtful reading of the manuscript, and for useful suggestions and comments which made for a clearer manuscript. This work was partially supported by funding from the Center for Exoplanets and Habitable Worlds. The Center for Exoplanets and Habitable Worlds is supported by the Pennsylvania State University, the Eberly College of Science, and the Pennsylvania Space Grant Consortium. This work was supported by NASA Headquarters under the NASA Earth and Space Science Fellowship Program through grants NNX16AO28H and 80NSSC18K1114. We acknowledge support from NSF grants AST-1006676, AST-1126413, AST-1310885, AST-1517592, AST-1310875, the NASA Astrobiology Institute (NAI; NNA09DA76A), and PSARC in our pursuit of precision radial velocities in the NIR. Computations for this research were performed on the Pennsylvania State University’s Institute for CyberScience Advanced CyberInfrastructure (ICS-ACI). We acknowledge support from the Heising-Simons Foundation via grant 2017-0494.

These results are based on observations obtained with the Habitable-zone Planet Finder Spectrograph on the Hobby-Eberly Telescope. We thank the Resident astronomers and Telescope Operators at the HET for the skillful execution of our observations of our observations with HPF. The Hobby-Eberly Telescope is a joint project of the University of Texas at Austin, the Pennsylvania State University, Ludwig-Maximilians-Universität München, and Georg-August Universität Gottingen. The HET is named in honor of its principal benefactors, William P. Hobby and Robert E. Eberly. The HET collaboration acknowledges the support and resources from the Texas Advanced Computing Center. 

These results are based on observations obtained with the 3\,m Shane Telescope at Lick Observatory. The authors thank the Shane telescope operators, AO operators, and laser operators for their assistance in obtaining these data. These results are based on observations obtained with the Apache Point Observatory 3.5-meter telescope which is owned and operated by the Astrophysical Research Consortium. We wish to thank the APO 3.5m telescope operators in their assistance in obtaining these data.

This paper includes data collected by the \textit{Kepler} telescope. The \textit{Kepler} and \textit{K2} data presented in this paper were obtained from the Mikulski Archive for Space Telescopes (MAST). Space Telescope Science Institute is operated by the Association of Universities for Research in Astronomy, Inc., under NASA contract NAS5-26555. Support for MAST for non-HST data is provided by the NASA Office of Space Science via grant NNX09AF08G and by other grants and contracts. Funding for the \textit{K2} Mission is provided by the NASA Science Mission directorate. This research made use of the NASA Exoplanet Archive, which is operated by the California Institute of Technology, under contract with the National Aeronautics and Space Administration under the Exoplanet Exploration Program. This work has made use of the VALD database, operated at Uppsala University, the Institute of Astronomy RAS in Moscow, and the University of Vienna. This work has made use of data from the European Space Agency (ESA) mission {\it Gaia} (\url{https://www.cosmos.esa.int/gaia}), processed by the {\it Gaia} Data Processing and Analysis Consortium (DPAC, \url{https://www.cosmos.esa.int/web/gaia/dpac/consortium}). Funding for the DPAC has been provided by national institutions, in particular the institutions participating in the {\it Gaia} Multilateral Agreement.

\facilities{\textit{K2}, \textit{Gaia}, ARCTIC/ARC 3.5m, ShaneAO/Lick 3m, HPF/HET 10m.} 
\software{AstroImageJ \citep{collins2017}, 
\texttt{astroplan} \citep{morris2018},
\texttt{astropy} \citep{astropy2013},
\texttt{astroquery} \citep{astroquery},
\texttt{barycorrpy} \citep{kanodia2018}, 
\texttt{batman} \citep{kreidberg2015},
\texttt{corner.py} \citep{dfm2016}, 
\texttt{celerite} \citep{Foreman-Mackey2017}, 
\texttt{emcee} \citep{dfm2013},
\texttt{everest} \citep{luger2017}, 
\texttt{EXOFASTv2} \citep{eastman2017},
\texttt{GNU parallel} \citep{Tange2011},
\texttt{iDiffuse} \citep{stefansson2018b},
\texttt{Jupyter} \citep{jupyter2016},
\texttt{juliet} \citep{Espinoza2018},
\texttt{matplotlib} \citep{hunter2007},
\texttt{numpy} \citep{vanderwalt2011},
\texttt{MC3} \citep{cubillos2017},
\texttt{MRExo} \citep{kanodia2019},
\texttt{pandas} \citep{pandas2010},
\texttt{pyde} \citep{pyde},
\texttt{radvel} \citep{fulton2018},
\texttt{SERVAL} \citep{zechmeister2018},
\texttt{statsmodels} \citep{statsmodels},
\texttt{telfit} \citep{gullikson2014},
\texttt{VESPA} \citep{morton2015vespa,morton2012vespa}.}

\newpage

\appendix

\section{HPF RV Drifts}
\begin{figure}[H]
\begin{center}
\includegraphics[width=0.7\textwidth]{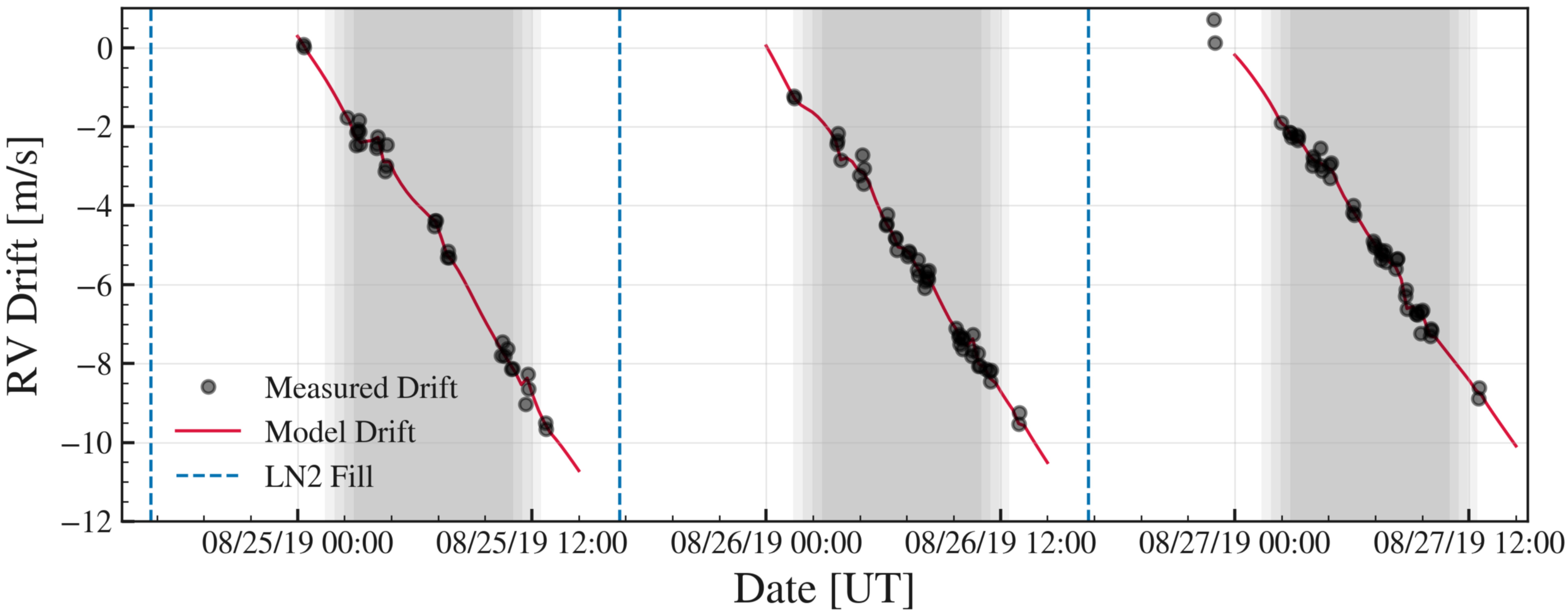}
\end{center}
\vspace{-0.5cm}
\caption{HPF RV drift as measured by the HPF LFC over three consecutive nights (grey shaded regions), to illustrate the characteristic saw-tooth drift behavior of HPF, which remains linear throughout a given night allowing us to precisely model and correct observations without simultanous LFC observations. The black points show the RV drift as measured by the HPF LFC through the HPF calibration fiber. The red curve shows the RV drift model (see discussion in text), and the blue dashed lines show the timing of the HPF LN2 fills.}
\label{fig:drift3}
\end{figure}

HPF has a characteristic sawtooth drift pattern at the $\sim$10-15m/s level throughout a 24 hour period (see Figure \ref{fig:drift3}). This drift pattern is related to the filling of the HPF liquid nitrogen (LN2) tank \citep{metcalf2019}, which is filled every morning to allow HPF to maintain a stable operating temperature. Although this drift characteristic of HPF has been discussed in detail in \cite{metcalf2019}, we further detail here the RV drift of HPF during the four nights we obtained RV observations of G 9-40, and discuss the contribution the RV drift has on the total estimated RV error.

\begin{figure}[H]
\begin{center}
\includegraphics[width=0.7\textwidth]{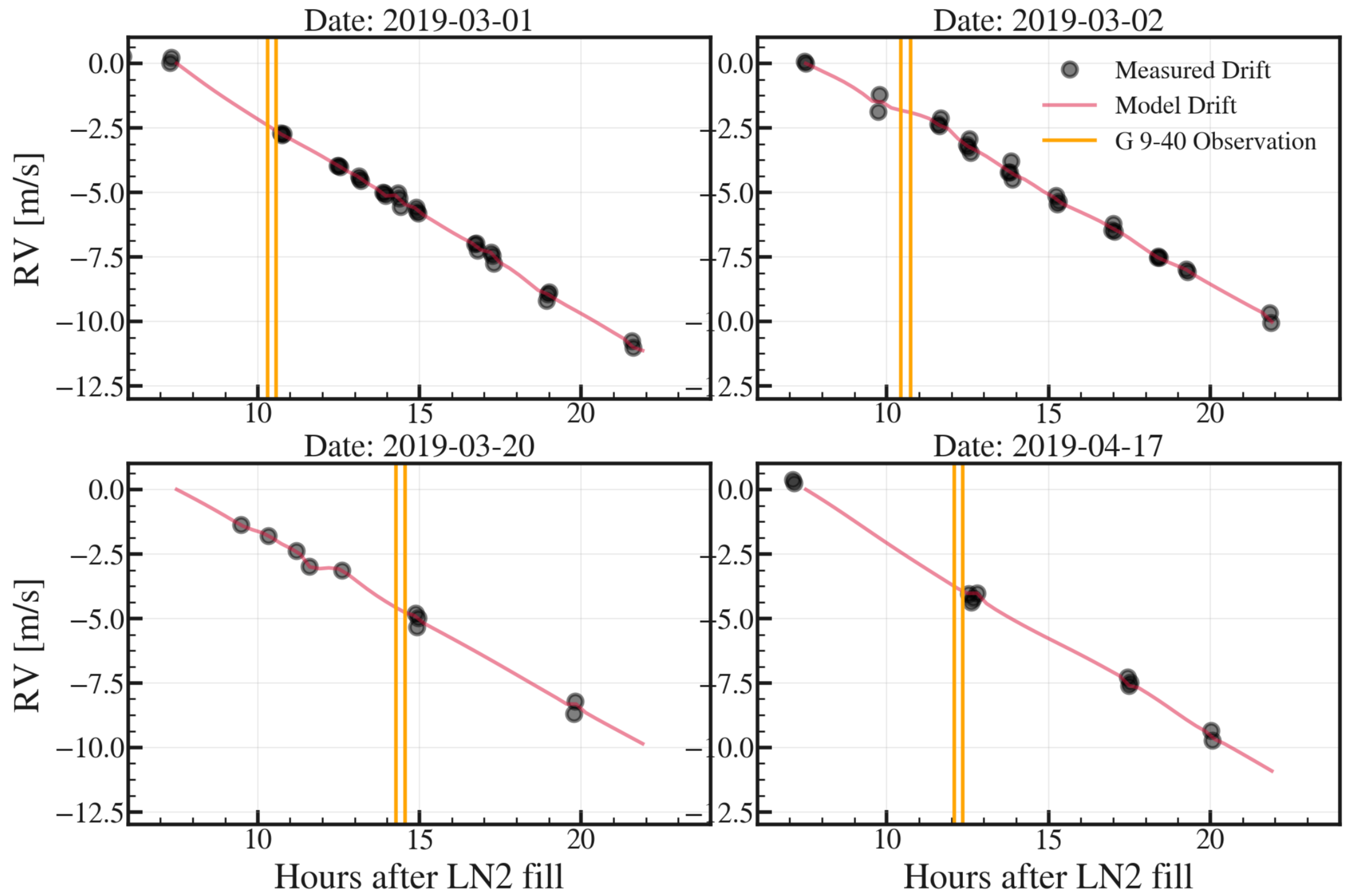}
\end{center}
\vspace{-0.5cm}
\caption{Nightly drift of HPF during the observations of G 9-40. The black points show when HPF LFC Calibration observations were obtained for the drift estimates used in this work. The red curve denotes the HPF model drift used to estimate HPF drifts throughout the night. The orange vertical lines show the times when G 9-40 observations were taken.}
\label{fig:drift}
\end{figure}

Figure \ref{fig:drift} shows the drift of HPF during the four nights we obtained G 9-40 RV observations. The black points in Figure \ref{fig:drift} show the RV drift derived from the HPF LFC through the HPF calibration fiber when no star was being observed, but such exposures are taken throughout the night to monitor the drift of HPF through the calibration fiber. The dedicated LFC drift-monitoring exposures all had a fixed exposure time of 106.5 s. The red curve in Figure \ref{fig:drift} shows the drift model used to estimate the drift of HPF between the black points. The timing of the 8 exposures of G 9-40 we obtained is denoted by the orange vertical lines in Figure \ref{fig:drift}. From Figure \ref{fig:drift} we see no major deviations from a linear drift behavior.

We calculate the HPF drift model (red curves in Figure \ref{fig:drift3} and \ref{fig:drift}) in three steps for a given night:
\begin{enumerate}
\item First, for a given night, the overall least-squares slope $s$ of the drift curve is estimated from the as-measured RV drifts $d_1,d_2,\ldots,d_N$, where $N$ is the number of HPF LFC calibration observations taken that night through the HPF calibration fiber.
\item Second, $N$ independent drift estimates $d_1^\prime,d_2^\prime,\ldots,d_N^\prime$ for the target observation are estimated by linearly extrapolating all of the individual LFC drift estimates $d_i$ to the flux-weighted midpoint of the target observation $T$, using the following formula,
\begin{equation}
d_i^\prime = s(T-t_i) + d_i,
\label{eq:delta}
\end{equation}
for all $i \in \{1,2,\ldots,N\}$.
\item Lastly, the final drift value $D(T)$ for our observation obtained at flux-weighted time $T$ is estimated by taking a weighted average of the $N$ independent drift values $d_1^\prime,d_2^\prime,\ldots,d_N^\prime$, where,
\begin{equation}
D = \frac{\sum_i^N w_i d_i^\prime}{\sum_i^N w_i},
\label{eq:drift}
\end{equation}
where the weights, $w_i = (T-t_i)^{-2}$, ensure that drift values closest in time to our observation carry the most weight in the average value.
\end{enumerate}

Since our drifts are dominated by systematics \citep{metcalf2019}, instead of photon noise, we estimate the error of each drift measurement empirically in two steps. First, we estimate the error on each LFC measured drift point using a leave-one-out approach. We drop the measurement whose error we want to estimate and then estimate the drift at that epoch using our drift model (Equations \ref{eq:delta} and \ref{eq:drift}). We then compute the difference between the estimated drift and our measured drift to obtain an estimate of the error of that single LFC measurement. Second, we treat these errors as independent and propagate this through the weighted average formula to estimate the drift at the epoch of G 9-40 observations. This leads to drift error estimates $<30 \unit{cm/s}$ for all of our G 9-40 observations, which we add in quadrature to our estimated RV errors. Our drift errors for each G9-40 observation are listed in Table \ref{tab:rvs}.

\bibliographystyle{yahapj}
\bibliography{references}

\end{document}